\documentclass[useAMS,usenatbib,onecolumn]{mnras}
\usepackage{amsmath,bm,amssymb}
\usepackage{dcolumn}
\usepackage[utf8]{inputenc}
\usepackage{graphics,graphicx}
\usepackage{float}
\usepackage{threeparttable}
\usepackage{url}
\usepackage{epstopdf}
\usepackage{siunitx}
\usepackage{epstopdf}
\usepackage{booktabs}

\usepackage{longtable}

\usepackage{color}

\newcommand{\cm}{cm$^{-1}$}

\newcommand{\Duo}{{\sc Duo}}

\newcommand{\ai}{\textit{ab initio}}
\newcommand{\Ai}{{\it Ab initio}}

\newcommand{\XS}{$X\,{}^{2}\Sigma^{+}$}
\newcommand{\AS}{$A\,{}^{2}\Pi$}
\newcommand{\BS}{$B\,{}^{2}\Sigma^{+}$}
\newcommand{\CS}{$C\,{}^{2}\Pi$}
\newcommand{\DS}{$D\,{}^{2}\Delta$}
\newcommand{\ES}{$E\,{}^{2}\Delta$}
\newcommand{\FS}{$F\,{}^{2}\Pi$}
\newcommand{\GS}{$G\,{}^{2}\Delta$}
\newcommand{\asi}{$a\,{}^{4}\Sigma^{+}$}
\newcommand{\bsi}{$b\,{}^{4}\Pi$}
\newcommand{\csi}{$c\,{}^{4}\Delta$}
\newcommand{\dsi}{$d\,{}^{4}\Sigma^{-}$}
\newcommand{\fsi}{$f\,{}^{4}\Sigma^{-}$}
\newcommand{\gsi}{$b\,{}^{4}\Pi$}

\newcommand{\SiNm}{$^{28}$Si$^{14}$N}

\newcommand{\name}{SiNful}

\graphicspath{{figures/}}

\title[ExoMol line lists -- {XLVI}: SiN]{ExoMol line lists -- {XLVI}: Empirical rovibronic spectra of silicon mononitrate (SiN) covering the 6 lowest electronic states and 4 isotopologues}

\date{\today}

\author[Semenov et al.]{Mikhail Semenov$^{1}$, Nicholas  Clark$^1$, Sergei  N. Yurchenko$^{1}$, Gap-Sue Kim$^2$, \newauthor{Jonathan Tennyson$^{1}$\thanks{The corresponding author: j.tennyson@ucl.ac.uk}}
\vspace*{4mm}\\
$^1$ Department of Physics and Astronomy, University College London, Gower Street, WC1E 6BT London, UK\\
$2$ Dharma College, Dongguk University, 30, Pildong-ro 1-gil, Jung-gu, Seoul 04620, Korea
}

\begin{document}

\label{firstpage}

\maketitle

\pagerange{\pageref{firstpage}--\pageref{lastpage}}

\begin{abstract}

Silicon mononitride ($^{28}$Si$^{14}$N, $^{29}$Si$^{14}$N, $^{30}$Si$^{14}$N, $^{28}$Si$^{15}$N) line lists covering infrared, visible and ultraviolet regions are presented. The \name\ line lists produced by ExoMol include rovibronic transitions between six electronic states:  \XS, \AS, \BS, \DS, \asi, \bsi. The  \ai\ potential energy and coupling curves, computed at the multireference configuration interaction (MRCI/aug-cc-pVQZ) level of theory, are refined for the observed states by fitting their analytical representations to 1052 experimentally derived SiN energy levels determined from rovibronic bands belonging to the $X$--$X$, $A$--$X$ and $B$--$X$ electronic systems through the MARVEL procedure. The SiNful line lists are compared to previously observed spectra, recorded and calculated lifetimes, and previously calculated partition functions. SiNful is available via the \url{www.exomol.com} database.  
\end{abstract}


\section{Introduction}

Silicon is considered to be the second most abundant element in Earths crust \citep{CRChandbook} and the seventh most abundant element on the planet according to \citet{95McSuxx.SiN}; low silicon abundance within a host is considered to be a better indicator of a potential planet detection than planet-metallicity correlation according to \citet{11BrDoCo.SiN}.
So far there have been multiple observations of SiN in different media in space: interstellar medium \citep{03ScLeMe.SiN, 77TuDaxx.SiN, 83FeMaBe.SiN}, red giant stars \citep{47DAxxxx.SiN,52GRxxxx.SiN}, and envelopes of carbon stars \citep{92TUxxxx.SiN}.
Apart from being an important astrophysical species, SiN has multiple different applications: quantum dot production \citep{17XiStRa.SiN}, quantum optomechanics \citep{18SeMoBo.SiN}, protective coating for biological tissues and dental implants \citep{13PeTkSc.SiN,SiNCoatingsDental}, membranes used in filtering and biosensor systems \citep{09VlApDm.SiN,14LeSoCh.SiN}. 

The SiN molecule seems to be well studied experimentally in comparison to some of its valence group counterparts. Originally the molecule was reported  by \citet{13Jexxxx.SiN}, with band heads attributed to the \BS--\XS\ band. Later \citet{25Muxxxx.SiN} conducted a more thorough study of the \BS--\XS\ reporting rotational transitions and detecting a weaker band later identified as \CS--\XS\ thanks to additional experimental works conducted by \citet{75Lixxxx.SiN}. Overall the main active electronic bands that have been rotationally resolved so far are \BS--\XS\ \citep{13OjGoxx.SiN,93ItSuKo.SiN,93NaCoMo.SiN,91PiCaxx.SiN,89FOxxxx.SiN,84WaAvDr.SiN,84FOxxxx.SiN,76BrDuHo.SiN,73SiBrRe.SiN, 69DuRaNa.SiN,68NaVexx.SiN,65ScBrxx.SiN,63StFexx.SiN,28Jedexx.SiN}, \AS--\XS\ \citep{04MeShFe.SiN,92ElHaGu.SiN, 89FOxxxx.SiN,88YaHiYa.SiN,85YaHixx.SiN,85FoLuAm.SiN,84FOxxxx.SiN,76BrDuHo.SiN,13Jexxxx.SiN} and \CS--\XS\ \citep{89FOxxxx.SiN,76BrDuHo.SiN,75Lixxxx.SiN, 25Muxxxx.SiN}, \BS--\AS\ \citep{76BrDuHo.SiN}. Of these \citet{13OjGoxx.SiN} is the most recent experimental study, which apart from the \BS--\XS\ band mentioned earlier, has also been able to detect and vibrationally resolve the \fsi--\dsi, \gsi--\dsi.  Additionally we want to highlight that \citet{68NaVexx.SiN} originally assigned their spectrum and quantum numbers to the SiO${}^+$ anion, however, that was later corrected by the \citet{69DuRaNa.SiN} and assigned to the \BS--\XS\ band of SiN.


A large number of theoretical investigations of SiN have been performed since the discovery of the molecule \citep{18XiShSu.SiN,13XiShSu.SiN,11OyPeWi.SiN, 09LiWaDi.SiN, 05KeMaxx.SiN, 00JuFrJa.SiN, 99SiSaBo.SiN, 98CaMaFr.SiN, 96Boxxxx.SiN, 94CHxxxx.SiN, 93ChKrFi.SiN, 92McLiCh.SiN, 91CuRaTr.SiN, 91MeHoxx.Sin, 89MuLaxx.SiN, 84ZiClSa.SiN, 84BrDoPe.SiN, 78PrBuPe.SiN,75GoShxx.SiN}. The most recent \ai\ work at the time of writing was carried out by \citet{18XiShSu.SiN} who conducted a thorough analysis of the transition dipole moment curves and potential energy curves for the lowest 8 doublets (\XS, \AS, \BS,  \CS, \DS, \ES,\FS, \GS), of SiN, additionally reporting on Frank-Condon factors and lifetimes for the first 15 vibrational levels of each electronic state, for which they  used  program LEVEL due to  \citet{98LeRoy.methods}. The states were studied at the Internally Contracted Multirefrence Configuration Interaction with Davidson correction (icMRCI+Q) level with additional extrapolation procedure employed as described by \citet{14OyKrKe.SiN}. A lot of work in the 2018 paper was built on the foundation of their earlier paper \citep{13XiShSu.SiN}, where they reported PECs for the  lowest 8 doublets (\XS, \AS, \BS, \CS, \DS, \ES, \FS, \GS) and 4 quartets (\asi, \bsi, \csi, \dsi ) and the lowest sextet state $1^6\Sigma^{+}$. Apart form PECs this paper also reported spectroscopic constants and discussed the effects of the spin-orbit couplings.

The work  described below will present original \ai\  potential energy curves (PECs), spin-orbit curves (SOCs), electronic angular momentum curves  (EAMCs) and (transition) dipole moment curves (DMCs)   calculated using a high level of theory with MOLPRO~\citep{Molpro:JCP:2020} and then empirically fitted using data obtained from a MARVEL  (measured active rotation vibration energy levels) \citep{jt412} procedure and generated from effective Hamiltonians using PGOPHER \citep{PGOPHER}. The refined spectroscopic model is then used to compute molecular line lists for four isotopologues of SiN. Both the refinement and line list calculations are performed using the rovibronic program \Duo\ \citep{Duo}. We then use our line lists to give high accuracy comparison to previously reported experimental spectra, calculated and experimental lifetimes, and partition function.

\section{\Ai\ model}
\label{sec:abinitio}

\begin{figure}
    \centering
    \includegraphics[width=0.7\textwidth]{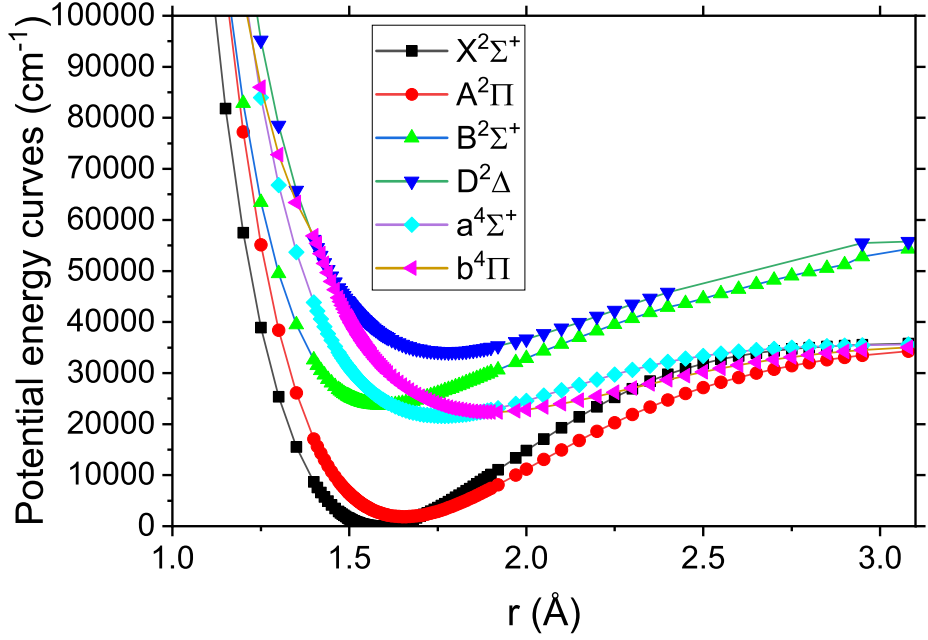}
    \caption{\Ai\ PECs calculated at the icMRCI level of theory using aug-cc-pVQZ basis set. The grid was specifically made denser around the equilibrium points of the \XS\ and \AS\ states.}
    \label{fig:abinitio_PEC}
\end{figure}

\begin{figure}
    \centering
    \includegraphics[width=0.45\textwidth]{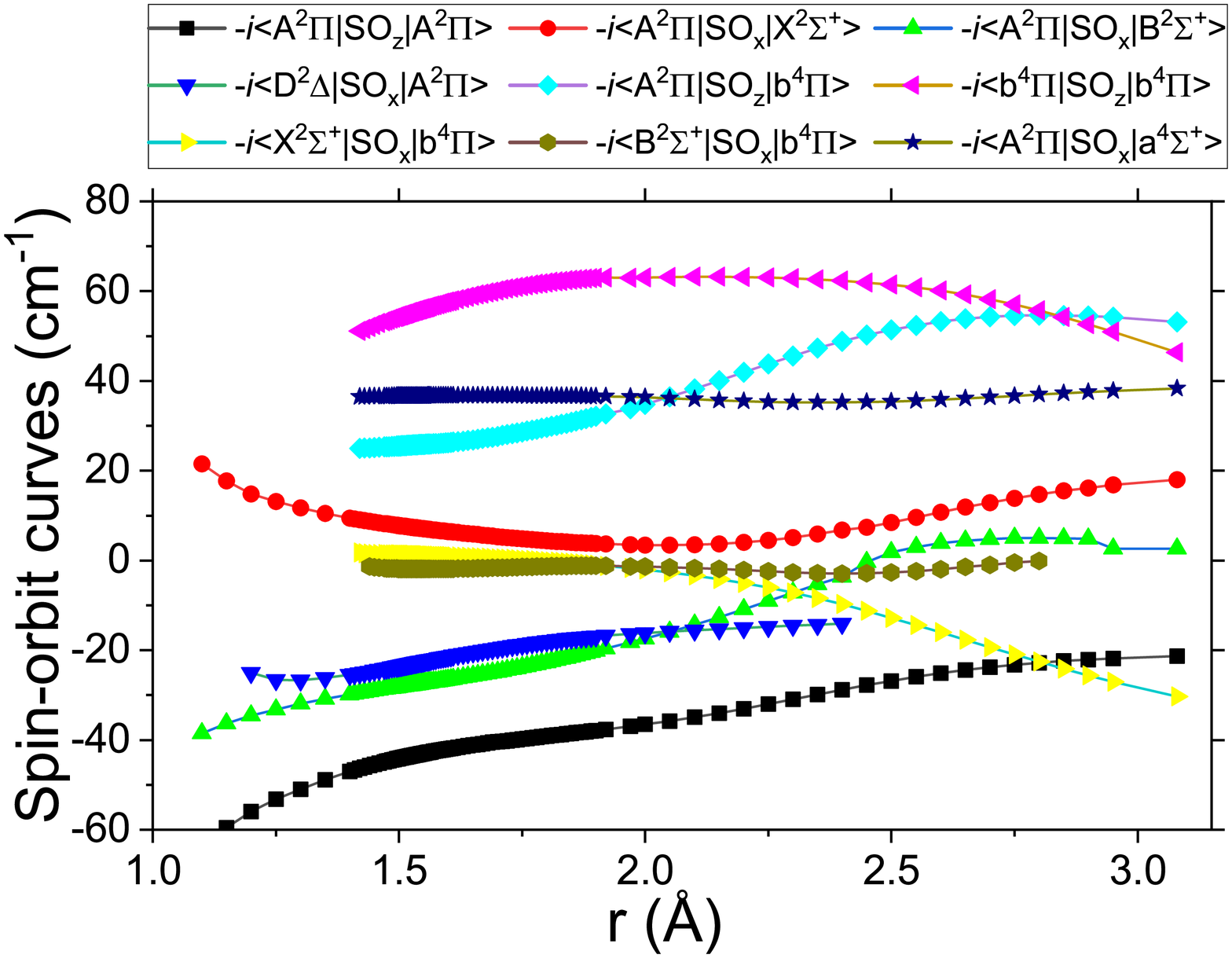}
        \includegraphics[width=0.45\textwidth]{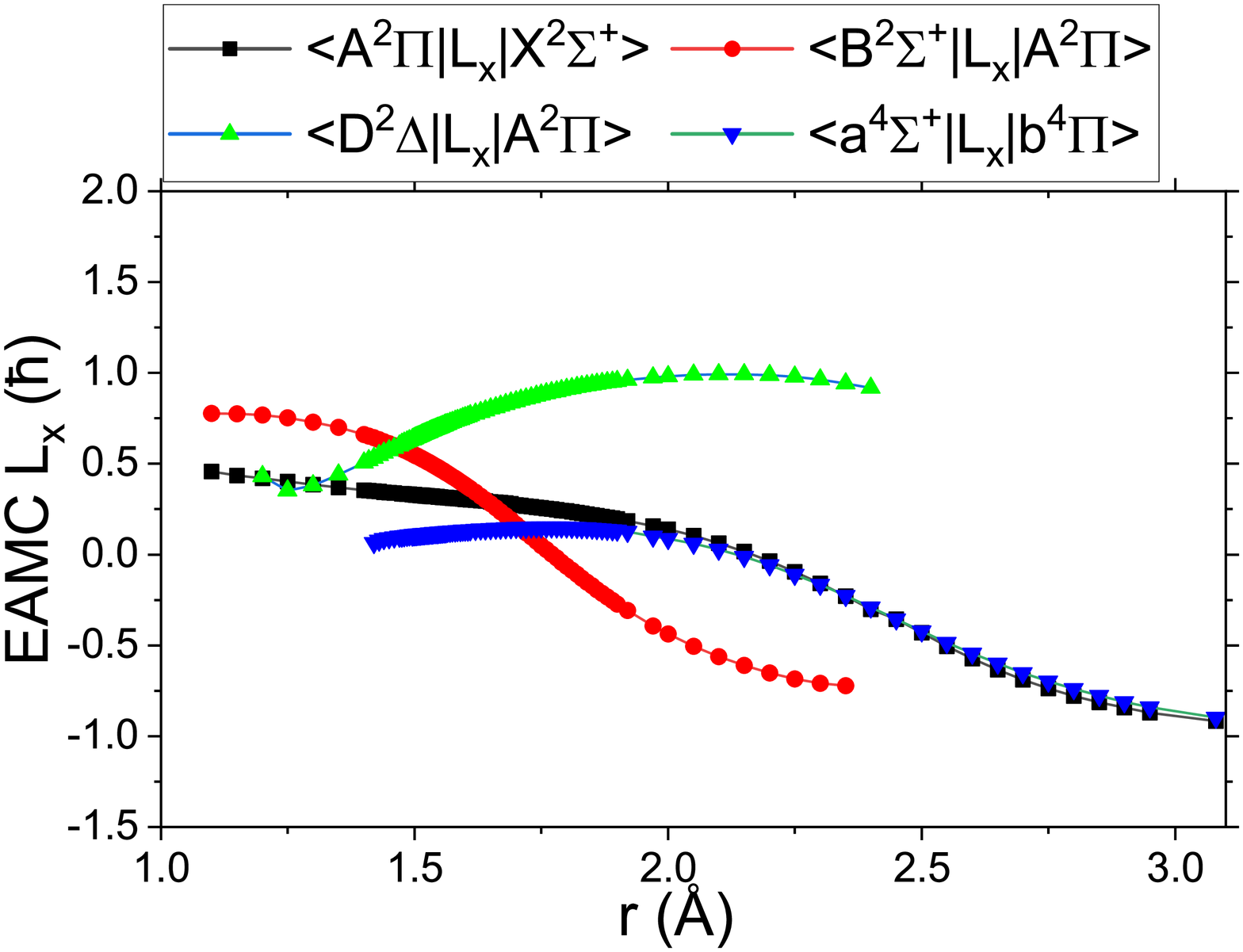}
    \caption{\Ai\ coupling curves of SiN at the icMRCI/aug-cc-pVQZ level of theory. Left: spin-orbit matrix elements $\langle i|$SO$|j \rangle$ for SiN at the icMRCI level of theory using aug-cc-pVQZ basis set. The MOLPRO values of the magnetic quantum numbers $M_S$ ($\Sigma$) for the curves can be found in Table.~\ref{t:mS_values}. Right: Electronic angular momentum $L_x$ curves.}
    \label{fig:abinitio_SOCs}
\end{figure}

\begin{figure}
    \centering
    \includegraphics[width=0.45\textwidth]{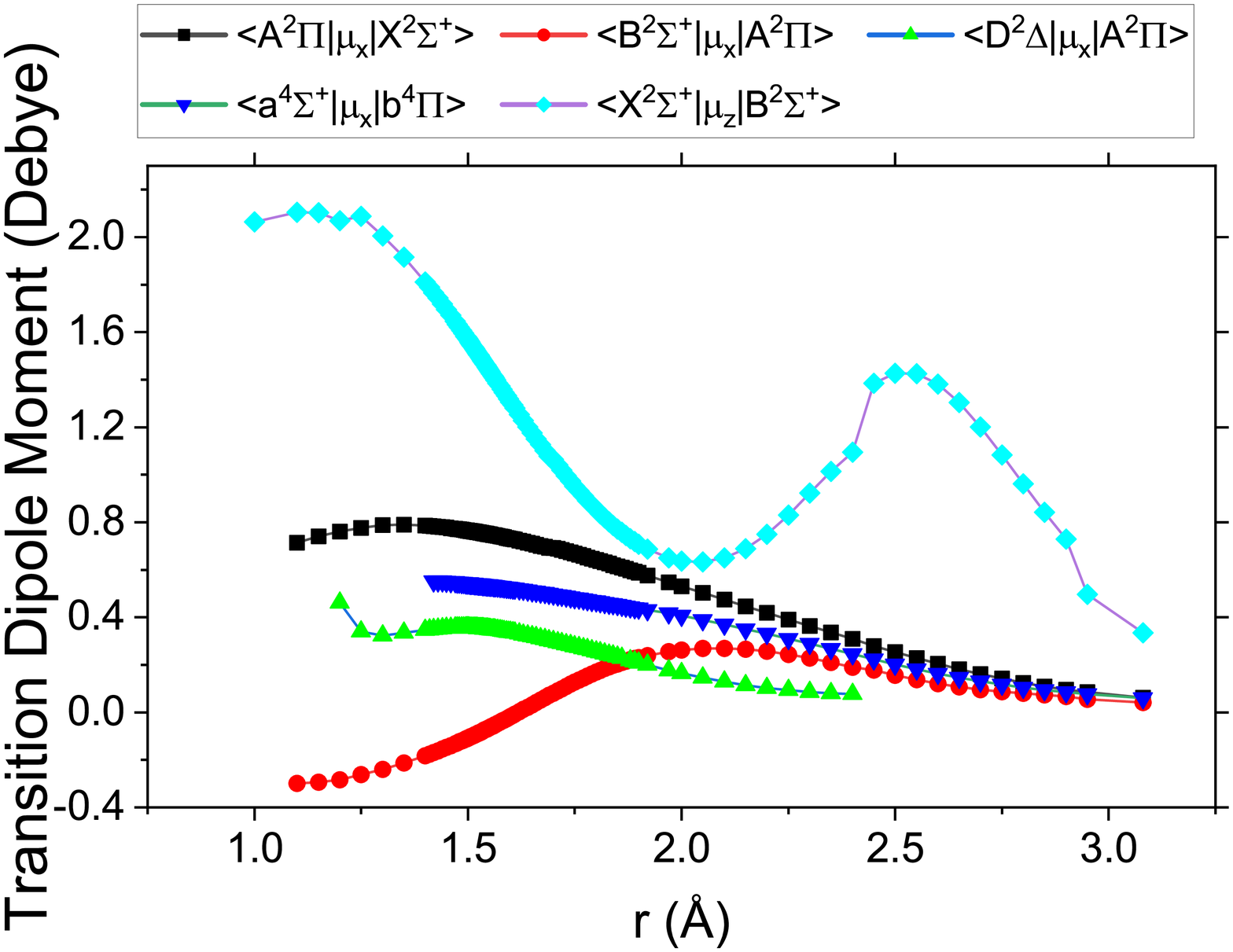}
    \includegraphics[width=0.45\textwidth]{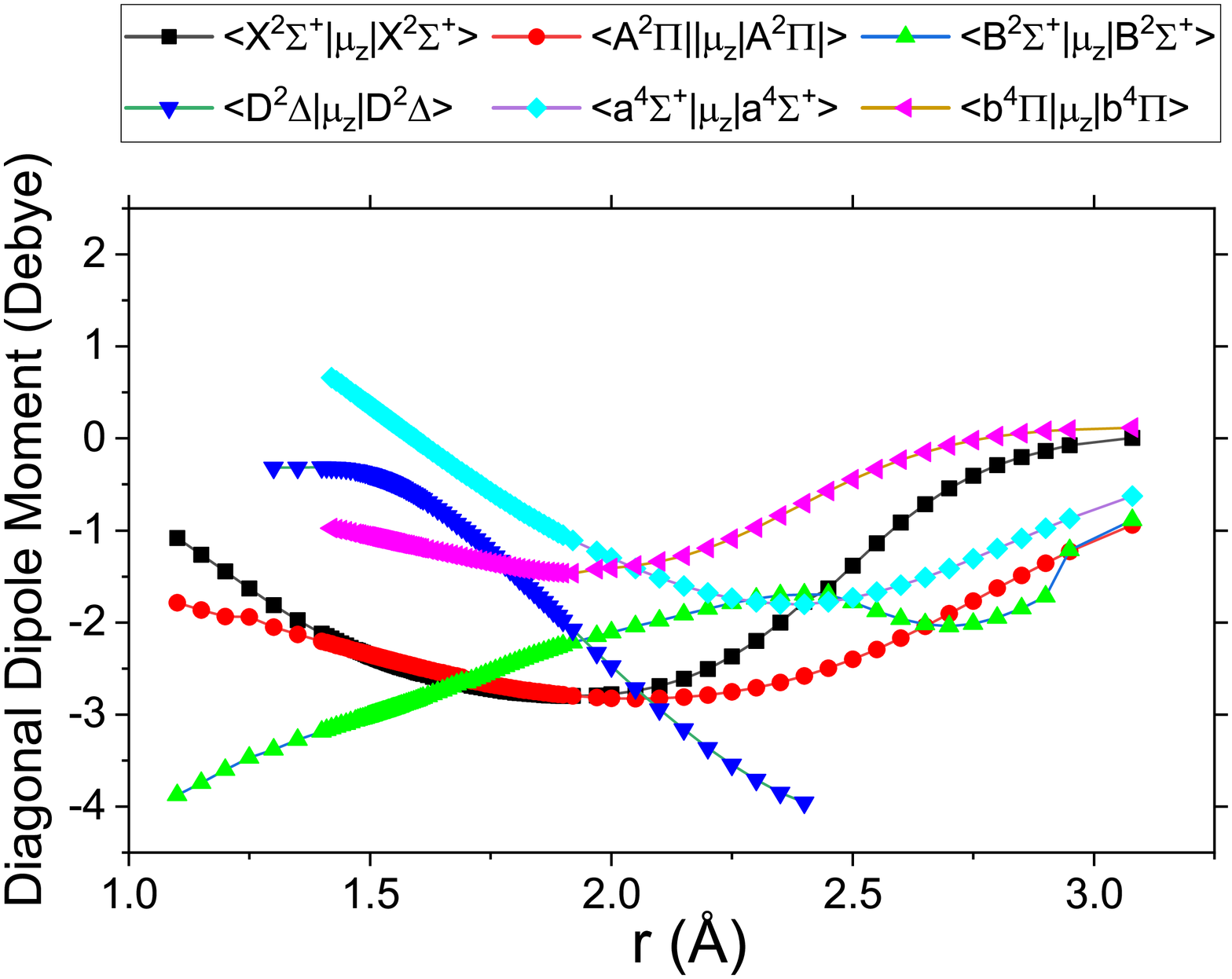}
    \caption{\Ai\ transitional and diagonal dipole moment curves calculated at the icMRCI level of theory using aug-cc-pVQZ basis set.}
    \label{fig:abinitio_DDM}
\end{figure}

In this work six lowest electronic states, namely \XS, \AS, \BS, {$D\,{}^{2}\Delta$}, \asi, \bsi, were studied using the state-averaged complete active space self consistent field (CASSCF) and multireference configuration interaction (MRCI) methods and  aug-cc-pVQZ basis sets.  The calculations were performed using  C$_{2v}$ point group symmetry. 
Potential energy curves, illustrated in Fig.~\ref{fig:abinitio_PEC}, spin-orbit coupling and  electronic angular moment (Fig.~\ref{fig:abinitio_SOCs}), dipole moment curves and transition dipole moment curves (Fig.~\ref{fig:abinitio_DDM}) were computed using MOLPRO2020 \citep{Molpro:JCP:2020}.
If the MOLPRO calculations at some geometries did not converge, they were interpolated or extrapolated from the neighbouring points as part of the rovibronic calculations (see below). We used an adaptive  \ai\ grid  consisting of 121 bond lengths ranging from 1.1 to 3.08~\AA\  with more points around the equilibrium region of the \XS\ state centred at 1.585 \AA, see Figs.~\ref{fig:abinitio_PEC}--\ref{fig:abinitio_DDM}, where the density of the \ai\ grid is shown.
The grid points with the corresponding  \ai\ values (if converged) are included in the supplementary material.


For \ai\ calculation we have considered the states of SiN correlating with two dissociation asymptotes, Si(\textsuperscript{3}P) + N(\textsuperscript{4}S) and Si(\textsuperscript{3}P) + N(\textsuperscript{2}D), which generate {$\,{}^{2}\Sigma^{+}$}(three states), {$\,{}^{2}\Pi$} (one state), {$\,{}^{4}\Sigma^{+}$}(one states) and {$\,{}^{4}\Pi$} (one state) which show six electronic states,  \XS, \AS, \BS, {$D\,{}^{2}\Delta$}, \asi\ and  \bsi\ which are mentioned above in \ai\ model section. 
Under the C$_{2v}$ point group symmetry,  $A\,{}_{1}$ corresponds  to ${}^{}\Sigma^{+}/{}^{}\Delta$. Thus, the {$\,{}^{2}\Delta$} state arose from the calculation of {$\,{}^{2}\Sigma^{+}$}. Molecular orbitals for the subsequent CI calculations were obtained for each spin and symmetry species from state averaged CASSCF calculations \citep{85WeKnxx.ai,85KnWexx.ai}  where the state averaging was achieved over all six electronic states considered in this work.

Within the $C_{2v}$ point group symmetry 14 (8{}$\sigma$, 3{}$\pi_{x}$,  3{}$\pi_{y}$) orbitals which contained 6 closed  (4{}$\sigma$, 1{}$\pi_{x}$,  1{}$\pi_{y}$) orbitals were used for all \ai\ calculations. Thus, the active space represents 8 active (4{}$\sigma$, 2{}$\pi_{x}$,  2{}$\pi_{y}$) orbitals with 9 active electrons and spans the valence orbitals $5\sigma$--$8\sigma$,$2\pi, 3\pi$. All CI calculations were carried out with the internally contracted CI method \citep{88WeKnxx.ai,88KnWexx.ai}. The
CI calculations used the same set of reference configurations used in the CASSCF calculation. 


The EAMCs represent the $L_x$ and $L_y$ matrix elements, which  are important for the accurate description of the lambda doubling effects in $\Pi$ states originating from the interactions with the $\Sigma$ and $\Delta$ states.
In Fig.~\ref{fig:abinitio_SOCs} (right) the $L_x$  components are shown, with $L_y$ related to them by symmetry. 

The definition of SOCs, which help form a complete and self-consistent description of a spectroscopic model, requires knowledge of the magnetic quantum numbers $M_S$ (i.e. the projection of the electronic spin $\Sigma$), which are specified in Table~\ref{t:mS_values}. 

All of the \ai\  curves are later mapped to a different, denser grid used in the solution of the rovibronic problem via interpolation and extrapolation as described in the Spectroscopic Model section below.

\begin{table}
    \caption{MOLPRO magnetic quantum numbers  $M_S$ values (also known as $\Sigma$, 
    projection of $\hat{\bf S}$  along the body fixed $z$ axis) for the \ai\ spin-orbit matrix elements  displayed in Fig.~\ref{fig:abinitio_SOCs}.  }
    \label{t:mS_values}
    \centering
    \begin{tabular}{r@{}c@{}lcc}
    \hline
    \multicolumn{3}{c}{SOC}& bra $M_S$ & ket $M_S$ \\
    \hline
    $\langle$ \AS$_y$ $\lvert$ & SO$_x$ & $\rvert$\XS$\rangle$ & -0.5 & 0.5  \\
    $\langle$ \AS$_y$ $\lvert$& SO$_x$& $\rvert$\BS$\rangle$ & -0.5 & 0.5  \\
    $\langle$ \AS$_y$ $\lvert$& SO$_x$&$\rvert$\asi$\rangle$ & -0.5 & 0.5  \\
    $\langle$ \DS$_{xy}$ $\lvert$& SO$_x$&$\rvert$\AS$_y\rangle$ & -0.5 & 0.5  \\
    $\langle$ \XS $\lvert$& SO$_x$&$\rvert$\bsi$_y\rangle$ & 0.5 &-0.5  \\
    $\langle$ \BS $\lvert$& SO$_x$&$\rvert$\bsi$_y\rangle$ & 0.5 &-0.5  \\
    $\langle$ \AS$_x$ $\lvert$& SO$_z$&$\rvert$\AS$_{x}\rangle$ & 0.5 & 0.5  \\
    $\langle$ \AS$_x$ $\lvert$& SO$_z$&$\rvert$\bsi$_x\rangle$ &  0.5 & 0.5   \\
    $\langle$ \bsi$_x$ $\lvert$& SO$_z$&$\rvert$\bsi$_x\rangle$ &  0.5 & 0.5   \\
    \hline
    \end{tabular}
 \end{table}

 The (transition) DMCs \AS--\XS, \BS--\XS, \BS--\AS, \DS--\AS, \asi--\bsi,\ and DMCs \XS--\XS, \AS--\AS, \BS--\BS, \DS--\DS, \asi--\asi\ and \bsi--\bsi\  were calculated \ai\ at the same level of theory as the PECs and are shown in Fig.~\ref{fig:abinitio_DDM}. 
 The phases of these non-diagonal TDMCs were selected to be consistent with the phases of the \ai\ curves produced in our MOLPRO calculations \citep{jt589}. There is a discontinuity in $\langle$\XS$|\mu_x|$\BS$\rangle$ at  2.5 \AA\ that we attribute to the interaction with the higher electronic states, which are not included in this model. Due to the large displacement from the equilibrium corresponding to  very high energies this discontinuity does not provide any material effect on our results, as can be seen from the spectra reproduced  below.

\section{MARVEL}

\label{sec:marvel}

All available experimental transition frequencies of SiN were extracted from the published spectroscopic literature and analysed using the MARVEL  procedure \citep{jt412,07CsCzFu.method,12FuCsxx.methods,jt750}. This procedure takes a set of assigned transition frequencies with measurement uncertainties and converts it into a consistent set of empirical energy levels with the uncertainties propagated from the input transitions. The transition data extracted as part of this work from the literature covers the three main bands of SiN involving the \XS, \AS, and \BS\ electronic states: \XS--\XS, \AS--\XS and \BS--\XS, as summarised in Table~\ref{tab:trans}. 

\subsection{Description of experimental sources}
All the available sources of experimental transitions considered in this work are listed below:

\textbf{92ElHaGu} by \citet{ 92ElHaGu.SiN}: an infrared (IR) study of the \AS--\XS\ system through 724 transitions. Unfortunately, the original line data with assignments were lost/unavailable and only the derived spectroscopic constants for \AS\ and \XS\ states remain. Using their \XS\ constants we produced
four \XS--\XS\ pseudo-experimental lines (MARVEL Magic numbers) to help connecting spectroscopic networks  in the MARVEL analysis. We also used their extended set of the spectroscopic constants in PGOPHER \citep{PGOPHER} to generate pseudo-experimental energies for the first five vibrational states of \AS\ and the first three vibrational states of \XS\ states (up to $J=49.5$) for the refinement of our spectroscopic model (see below). \citet{ 92ElHaGu.SiN} is the only existing information on the vibrationally excited \AS\ states of SiN and  is therefore crucial for providing MARVEL energies for states of  \AS\ with $v>1$. The PGOPHER file used to generate the energies for fitting  is  provided as part of the supplementary information. 

\textbf{85FoLuAm} by \citet{85FoLuAm.SiN}: This IR observation reported 187 lines of the \AS--\XS\ system originally assigned to the (2,0) band and 6 microwave transitions in the \XS--\XS\ system. 
This band was later reassigned to  1--0 by  \citet{88YaHiYa.SiN}. 

\textbf{85YaHiXX} by \citet{85YaHixx.SiN}: This IR study reported  170 lines of the \AS--\XS\ system originally assigned to the (1,0) band, which however  was later reassigned to (0,0) by  \citet{88YaHiYa.SiN}.

\textbf{76BrDuHo} by \citet{76BrDuHo.SiN}: This is a UV study of the \BS--\XS\ system. Only spectroscopic constants were reported, which we used to generate four pseudo-experimental \BS--\XS\ lines using PGOPHER to help connect the MARVEL inputs into a single spectroscopic network   with $J'=1.5$~\BS\  $\longleftarrow$ $J"=0.5$ \AS\  for the (1,0) and (2,0) bands ($e$ and $f$).

\textbf{75Linton} by \citet{75Lixxxx.SiN}: This UV study reported 460 lines of the \CS--\AS\ system. Sadly due to the low quality of this data and the lack of information provided in the paper to reconstruct the full set of quantum numbers required for MARVEL, this source is omitted from the MARVEL analysis of the  current work, but the original lines scanned using OCR (optical character reader) software are still provided as part of the supplementary information.

\textbf{68NaVe} by \citet{68NaVexx.SiN}: This UV study reported 422 lines of  the \BS--\XS\ system  covering the (0,0), (1,1), (2,2) bands. Originally the spectra was assigned to the SiO$^+$ molecules, but that was reassigned to  SiN later as per the correction in \citet{69DuRaNa.SiN}.

\textbf{29JeDe} by \citet{28Jedexx.SiN}: This 1929 work is still the most extensive to date UV observation of the \BS--\XS\ system of SiN  reporting 1355 lines from the  (0,0), (1,1), (2,2), (3,3), (4,4), (3,2), (4,3), (5,4), (6,5) vibronic bands.

In total, 1987 experimental and 9 pseudo-experimental transitions were processed via the online MARVEL app (available through a user-friendly web interface at \href{http://kkrk.chem.elte.hu/marvelonline}{http://kkrk.chem.elte.hu/marvelonline}) using the Cholesky (analytic) approach with a {0.5}~cm$^{-1}$ threshold on the uncertainty of the ``very bad'' lines. The final MARVEL process for \SiNm\ resulted in one main spectroscopic network, containing 1054 energy levels and  1456 validated transitions, with  the rotational excitation up to $J=44.5$ and covering energies  up to 30\,308~cm$^{-1}$. These energy levels in conjunction with energies generated from PGOPHER were used to refine our \ai\ rovibronic spectroscopic model (PECs, SOCs and EAMCs) presented above. The MARVEL input transitions and output energy files are given as part of the supplementary material. 


\begin{table*}
    \centering
    
    \caption{Breakdown of the assigned transitions by electronic bands for the sources used in this MARVEL study. {A} and {V}  are the numbers of the available and validated transitions, respectively. The mean and maximum uncertainties (Unc.) obtained using the MARVEL procedure are given in \cm.}
    \label{tab:trans}
    \resizebox{\textwidth}{!}{
    \begin{tabular}{llllcc}
     \toprule
     Electronic Band & Vibrational Bands & $J$ Range & A/V &   WN range \cm\ & Unc. (Mean/Max)  \\
     \midrule
     \textbf{29JeDeXX}\\                           
     \BS--\XS & (0,0),(1,1),(2,2),(3,2),(3,3) & 0.5--43.5 & 1060/1355 & 23267--24279 & 0.100/0.197 \\
     & (4,3),(4,4),(5,4),(5,5),(6,5) & 0.5--43.5 &  &  \\
     \textbf{68NaVexx}\\
     \BS--\XS & (0,0),(1,1),(2,2) & 0.5--44.5 & 420/422.0 & 23876--24280 & 0.030/0.191 \\
     \textbf{76BrDuHo}\\
     \BS--\XS & (1,0),(2,0) & 0.5--1.5 & 4/4 & 25236--26203 & 0.100/0.100 \\
     \textbf{84YaHiXX}\\                                 
     \AS--\XS & (0,0) & 0.5--31.5 & 195/195 & 1918--2037 & 0.007/0.007 \\
     \textbf{85FoLuAm}                   
     \AS--\XS & (1,0) & 0.5--36.5 & 199/199 & 2917--3039 & 0.001/1.00$\times10^{-3}$ \\
     \XS--\XS & (0,0) & 0.5--4.5 & 6/6 & 2.9039--5.8329 & 1.00$\times10^{-3}$/1.00$\times10^{-3}$ \\
     \textbf{92ElHaGu}
     \XS\--\XS\ & (1,0),(2,0) & 0.5--1.5 & 4/4 & 1140--2267 & 0.100/0.100 \\
     \bottomrule
     \end{tabular}
     }
\end{table*}


\section{Rovibronic calculations}

To obtain rovibronic energies and wavefunctions for the six electronic states in question, we solve a set of fully coupled Schr\"{o}dinger equation for the motion of nuclei using the \textsc{Duo} program \citep{Duo}. \Duo\ uses the Hunds case~a basis set with the vibrational basis functions obtained by solving uncoupled vibrational Schr\"{o}dinger equations for each electronic state in question with using a sinc DVR basis set \citep{82GuRoxx}. Atomic masses are used to represent the kinetic energy operator.   In \Duo\ calculations,  an equidistant grid of 501 radial points ranging from 1.1 to 5.0 \AA. The \ai\ curves were cubic-spline interpolated to map the \ai\ curves onto the denser \textsc{Duo} grid. For the  extension outside the \ai\ bond lengths ($r>3.080$ ~\AA), the \ai\ curves were extrapolated using the functional forms given by \citep{Duo}
\begin{eqnarray}
\nonumber
  f_{\rm PEC}^{\rm short}(r) &=& A + B/r, \\
  f_{\rm TDMC}^{\rm short}(r) &=& A r + B r^2,\\
  \nonumber
  f_{\rm other}^{\rm short}(r) &=& A + B r,
\end{eqnarray}
for short range and
\begin{eqnarray}
\nonumber
 f_{\rm PEC}^{\rm long}(r)  &=& A + B/r^6\\
 f_{\rm EAMC}^{\rm long}(r)  &=& A + B r, \\
 \nonumber
 f_{\rm other}^{\rm long}(r)  &=& A/r^2 + B/r^3
\end{eqnarray}
for  long range, where $A$ and $B$ are stitching parameters.
A more detailed description of the \textsc{Duo} methodology was previously given by \citet{Duo}, see also \citet{jt632}.


\subsection{Refining the spectroscopic model}


For our spectroscopic model of SiN we initially used \ai\  PECs for the doublets  \XS, \AS, \BS, \DS\ and quartets \bsi, \asi, as well as all appropriate SOCs and EAMCs of SiN. The \XS, \AS\ and \BS PECs and the associated couplings were then refined by fitting to our empirical set of MARVEL and PGOPHER term values of \SiNm\ as described above.

\begin{figure}
    \centering
    \includegraphics[width=0.80\textwidth]{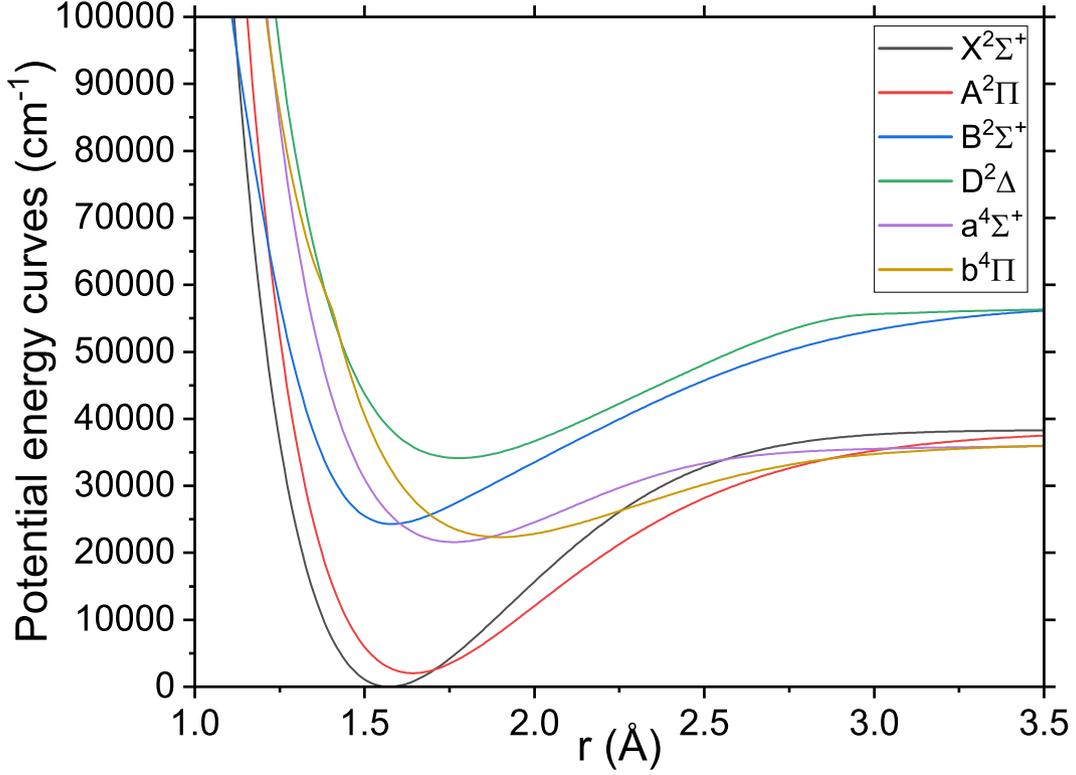}
    \caption{Potential energy curves of SiN representing our final spectroscopic and used in the  \textsc{Duo} calculations: \XS, \AS\ and \BS\ are refined while \DS, \asi\ and \bsi\ are \ai.}
    \label{fig:PECs}
\end{figure}

For the refinements, the PECs for the \XS, \AS,  and \BS\ states were parameterised using the  Extended Morse Oscillator (EMO) function \citep{EMO} as given by
\begin{equation}\label{e:EMO}
V(r)=V_{\rm e}\;\;+\;\;(A_{\rm e}-V_{\rm
e})\left[1\;\;-\;\;\exp\left(-\sum_{k=0}^{N} B_{k}\xi_p^{k}(r-r_{\rm e})
\right)\right]^2,
\end{equation}
where $A_{\rm e} - V_{\rm e} = D_{\rm e}$ is the dissociation energy, $A_{\rm e}$ is the corresponding asymptote, $r_{\rm e}$ is an equilibrium distance, and $\xi_p$ is the \v{S}urkus variable \citep{84SuRaBo.method} given by
\begin{equation}
\label{e:surkus:2}
\xi_p= \frac{r^{p}-r^{p}_{\rm e}}{r^{p}+r^{p}_{\rm e }}
\end{equation}
with $V_{\rm e} = 0$ for the \XS\ state. The \XS\ and \AS\ states have a common asymptote which was fixed for the analytical form of the potential to the ground state dissociation energy $D_{\rm e}$ 4.75~eV  based on the experiment by  \citet{93NaCoMo.SiN} (see also \citet{05KeMaxx.SiN}). Similarly the \BS\ state was adjusted to be 19233 \cm{} above the \XS\ dissociation using the atomic excitation energy of N \citep{NISTWebsite}.
The processed and refined PECs used in \textsc{Duo} can be seen in Fig.~\ref{fig:PECs}.

\begin{figure}
    \centering
    \includegraphics[width=0.44\textwidth]{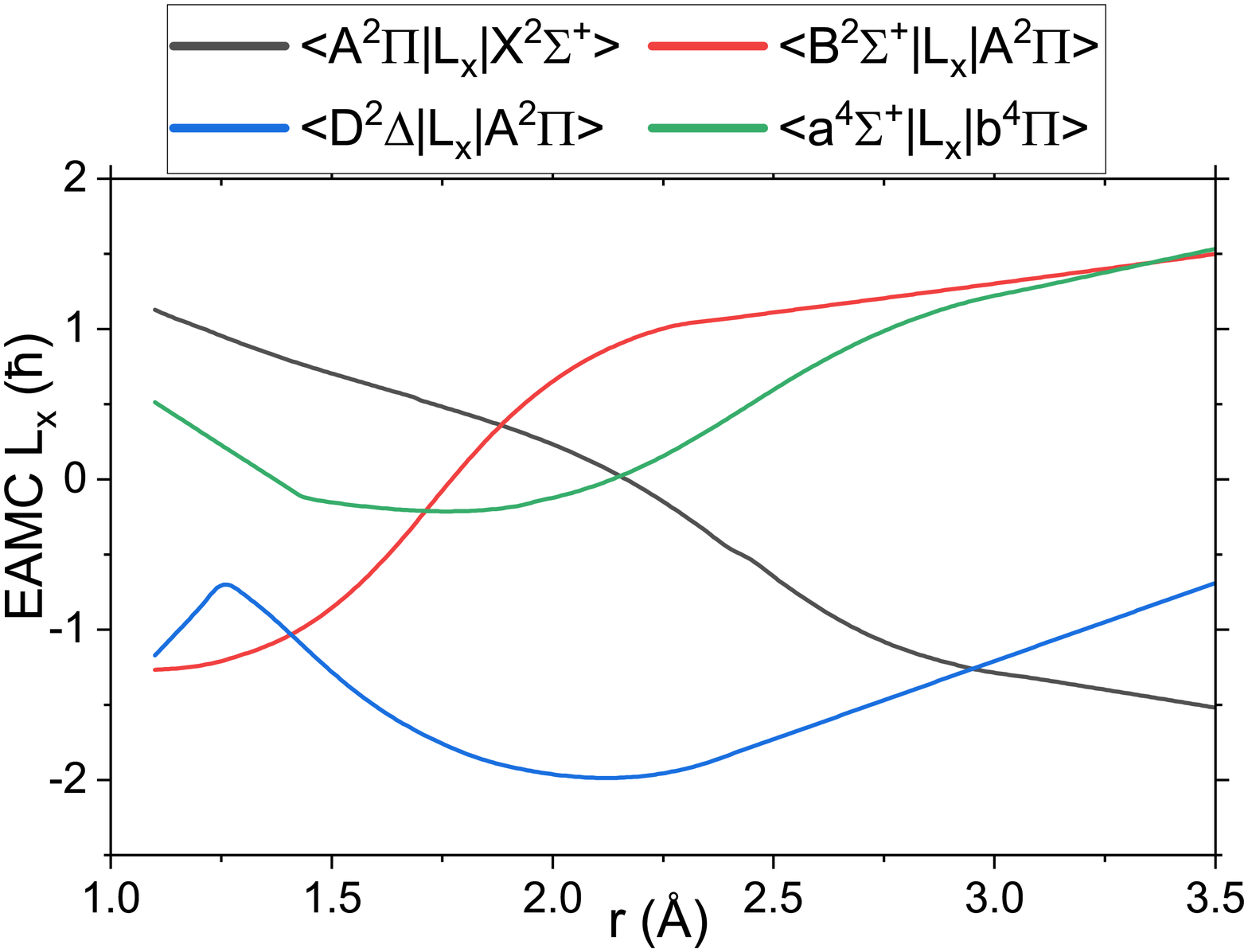}
    \includegraphics[width=0.44\textwidth]{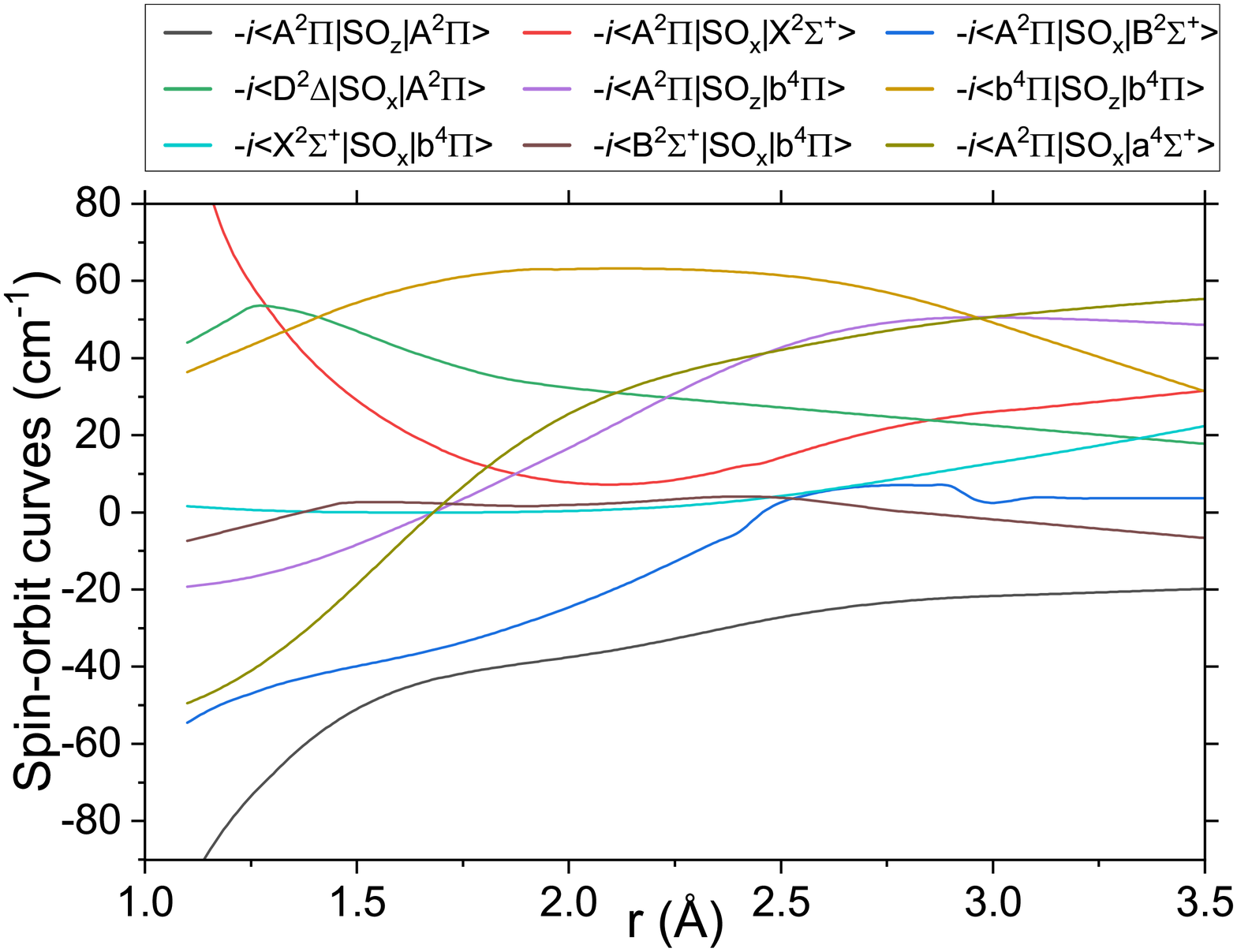}    
    \caption{Refined electronic angular momentum and spin-orbit curves for SiN. The MOLPRO values of the magnetic quantum numbers $M_S$ for the SOCs can be found in Table.~\ref{t:mS_values}}
    \label{fig:EAMC:SO}
\end{figure}

For the parameterisation of  SOCs and the EAMCs, the \ai\ curves were morphed using a polynomial decay expansion as given by:
\begin{equation}
\label{e:bob}
F(r)=\sum^{N}_{k=0}B_{k}\, z^{k} (1-\xi_p) + \xi_p\, B_{\infty},
\end{equation}
where $z$ is the damped-coordinate polynomial given by:
\begin{equation}\label{e:damp}
z = (r-r_{\rm ref})\, e^{-\beta_2 (r-r_{\rm ref})^2-\beta_4 (r - r_{\rm ref})^4} .
\end{equation}
The refined curves $f(r)$ are the represented as 
$$
f(r) = F(R) f^{\rm ai}(r), 
$$
with $B_{\infty} = 1$ in order for $F(r) \to 1$ at $r\to \infty$. Morphing allows one to retain the original shape of the property with a minimum number of varied parameters, see e.g.  \citet{jt703} and \citet{jt711}. In Eq.~\eqref{e:damp}, $r_{\rm ref}$ is a reference position chosen to be close to $r_{\rm e}$ of \XS\ and $\beta_2$ and $\beta_4$ are damping factors, typically chosen to be $8\times10^{-1}$ and $2\times10^{-2}$.
The morphed and extended by \textsc{Duo} SOCs and EAMCs are shown in Fig.~\ref{fig:EAMC:SO}.

\begin{figure}
    \centering
    \includegraphics[width=0.32\textwidth]{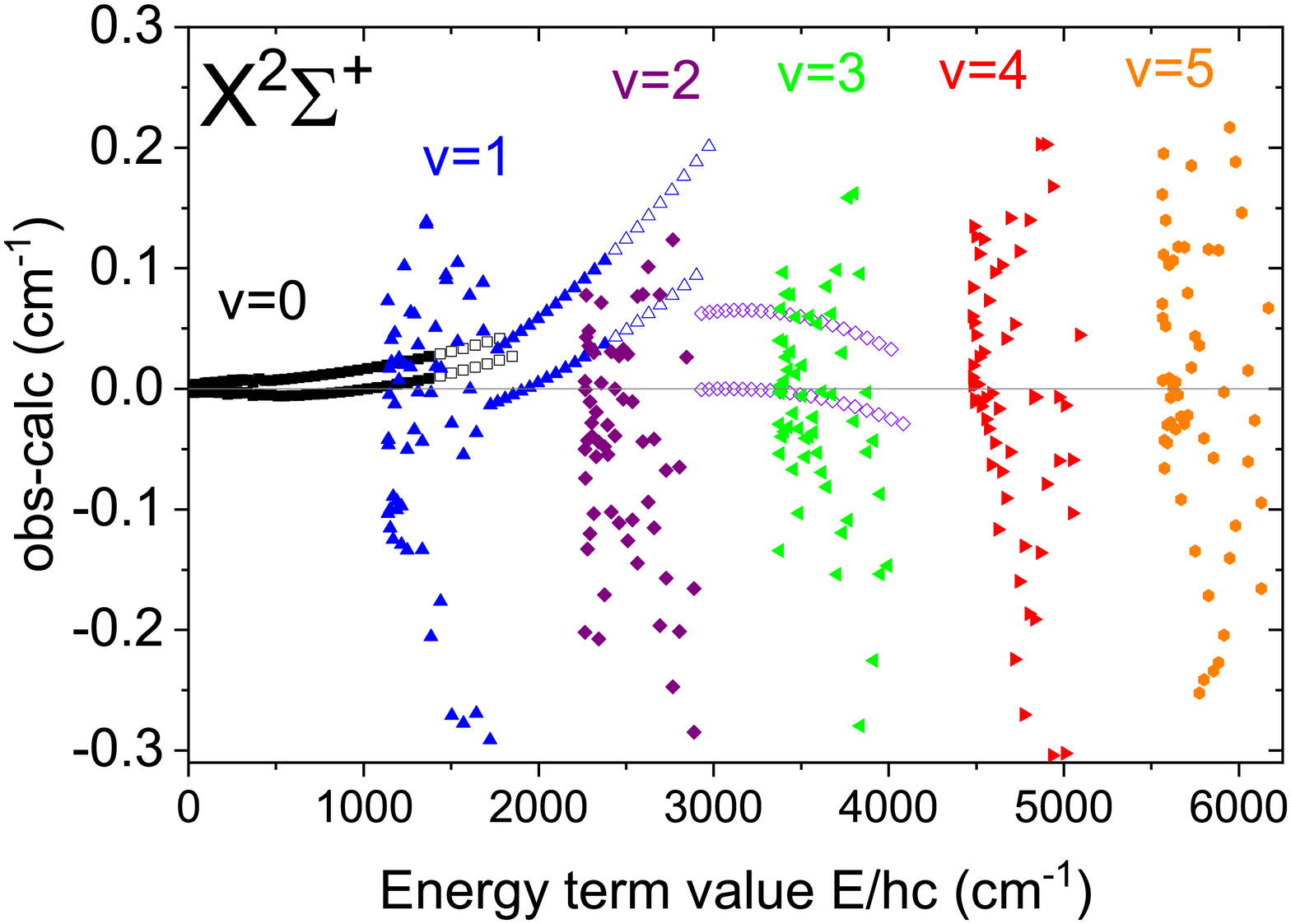}
    \includegraphics[width=0.33\textwidth]{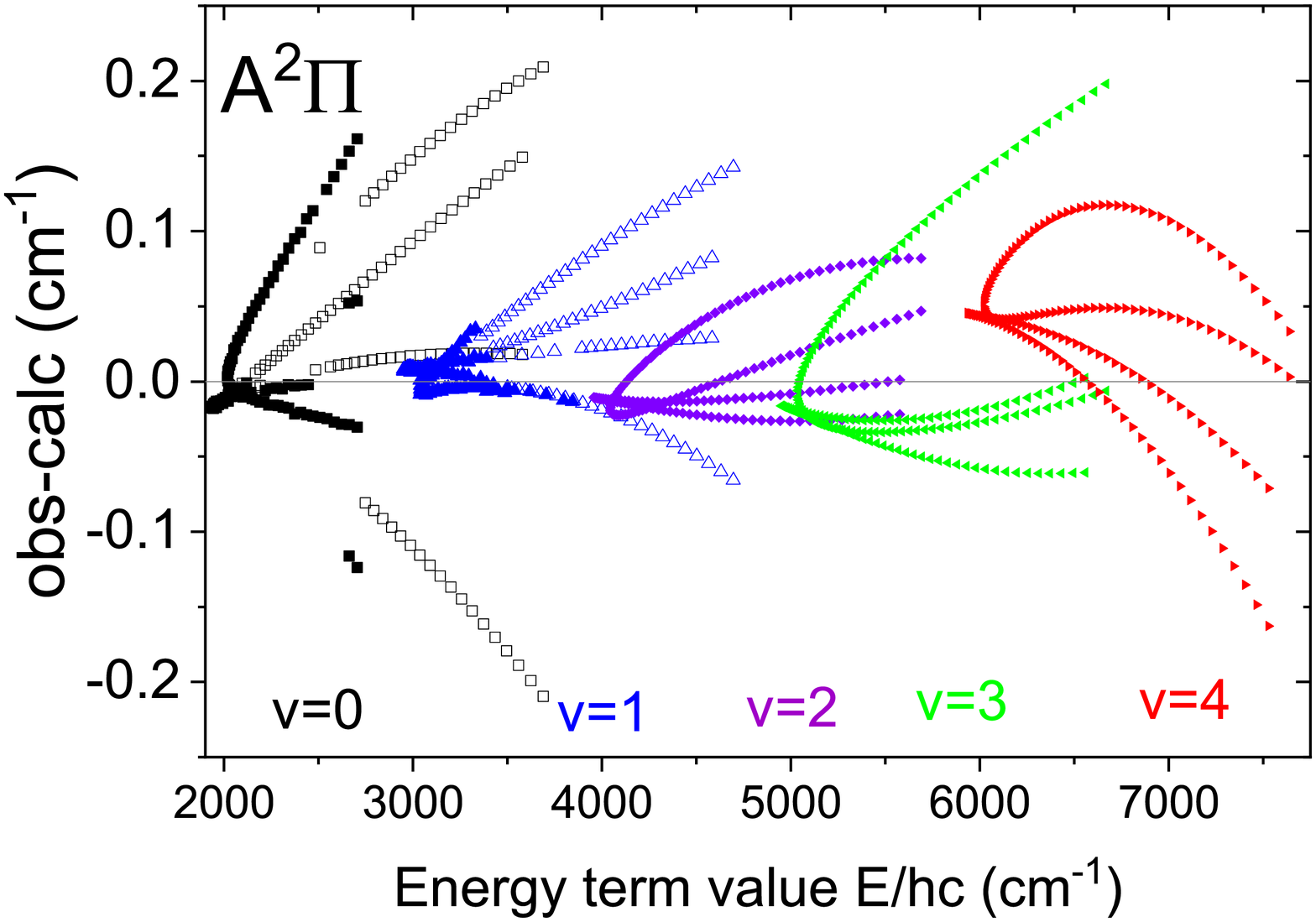}
    \includegraphics[width=0.33\textwidth]{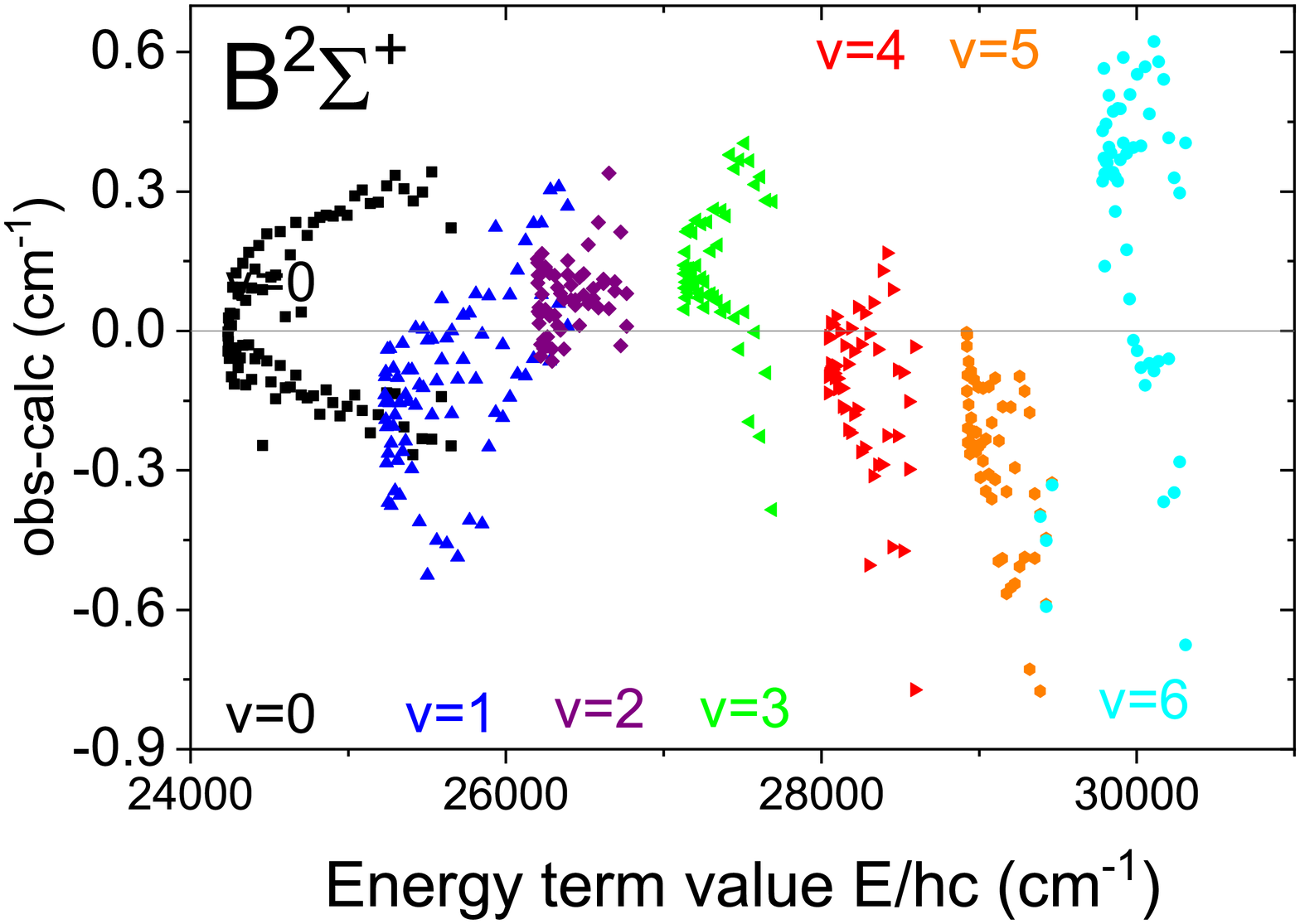}
    \caption{The residuals (Obs.-Calc.) between the experimentally determined energies of SiN (\XS, \AS\ and \BS)  from our MARVEL analysis (open), pseudo-experimental PGOPHER (filled) generated energies and  \textsc{Duo} energies corresponding to our refined spectroscopic model. Only MARVEL energies are available for \BS, hence the vibrational level labels are not differentiated by source. }
    \label{fig:residueAE}
\end{figure}

The 1062 MARVEL and 854 PGOPHER energy levels values were used to refine the PECs, SOCs, EAMCs into analytical form as described above. Figure~\ref{fig:residueAE} shows the residuals representing how well our model compares to the MARVEL and PGOPHER energies. Unfortunately the most significant work from the number of transitions provided is \citet{28Jedexx.SiN}, where the accuracy of the transitions was 0.1~cm$^{-1}$  at best. This is the main reason for the spread of uncertainties in the \XS\ and \BS\ states. Similarly most data for the \AS\ state had to be supplemented using the PGOPHER calculations due to the difficulty of obtaining experimental data for the state.


\subsection{Dipole moment curves}

\begin{figure}
    \centering
    \includegraphics[width=0.44\textwidth]{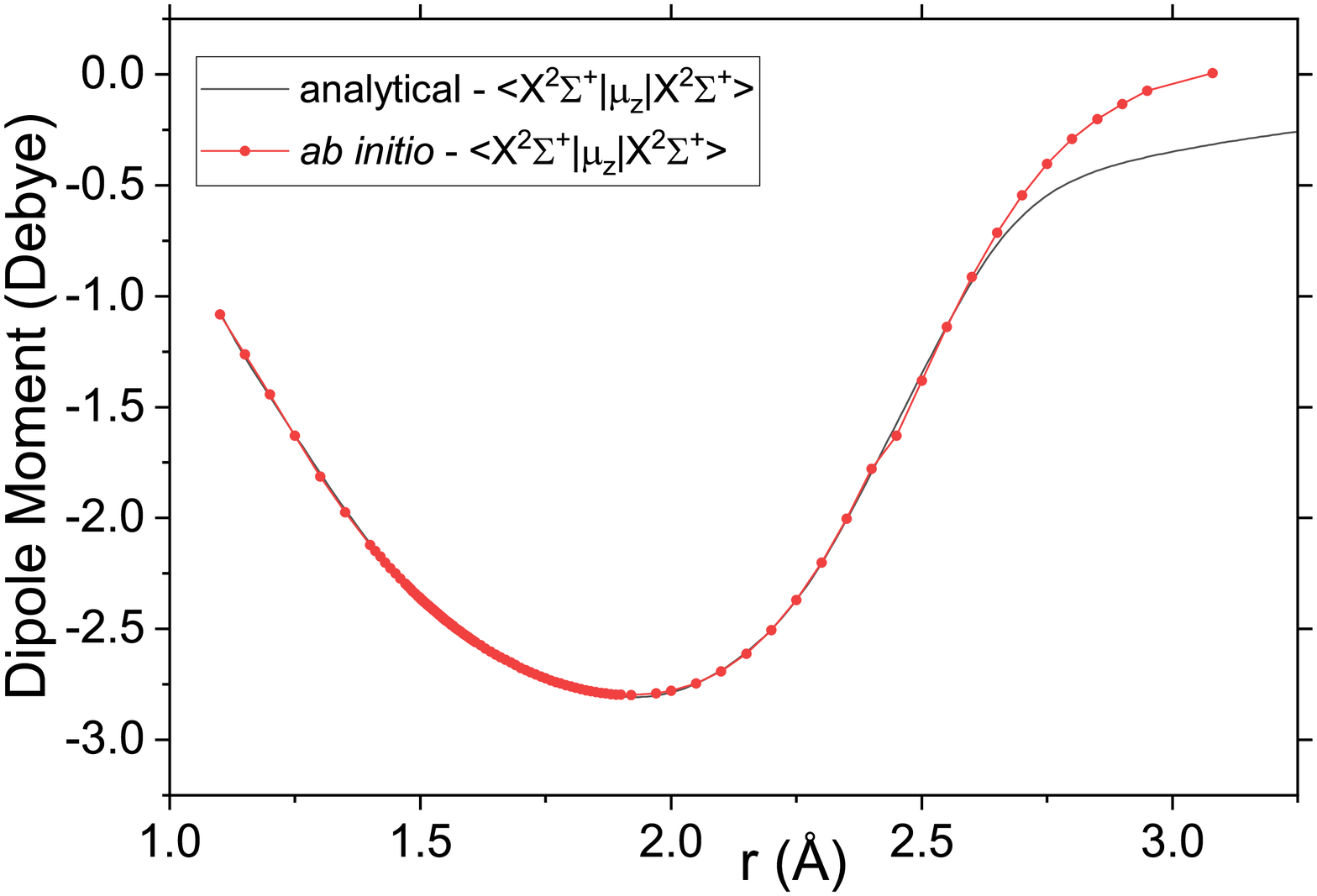}
    \caption{Comparison of the \ai\ and fitted DMC for the \XS--\XS\ system.}
    \label{fig:DMC_X_Comparison}
\end{figure}

Most of the (transition) dipole moment curves from Fig.~\ref{fig:abinitio_DDM} were left unchanged and used as grid points in \Duo\ calculations 
apart from the $\langle $\XS$|\mu_z|$\XS$\rangle$, which was fitted to analytical form  of the  polynomial decay given in Eq.~\eqref{e:bob}. Having an analytical form helps to reduce the numerical noise arising due to interpolation of \ai\ curves present in high overtone transitions, see \citet{16MeMeSt}. The comparison between \ai\ \XS--\XS\ DMC and its analytical form is shown in Fig.~\ref{fig:DMC_X_Comparison}, with the deviations at very large radial displacement not affecting the intensities for the energy excitations selected for this study.

All expansion parameters or curves defining our spectroscopic model are given as supplementary material to the paper as a \textsc{Duo} input file. 

\section{Line list and simulations of spectra of SiN}
\label{s:linelist}

The \name\ line list  was produced with \textsc{Duo} using the empirically refined and \ai\ curves as described above. For the main isotopologue \SiNm, it contains 43~646~806 transitions and 131~935 states for  \XS, \AS, \BS, \DS, \asi and \bsi,  covering wavenumbers up to 58~000 \cm\,  $v = 0 \ldots 30$  and $J = 0 \ldots  245.5$. For the isotopologue line lists only the atomic masses were adjusted in the \textsc{Duo} input files. Further details on the line list statistics covering the isotpologues can be seen in Table~\ref{t:iso}. The line list is provided in State  and Transition files, as is customary for the ExoMol format \citep{jt810}.  Extracts from the States and Trans files are shown  in Tables \ref{t:states} and \ref{t:trans}, respectively; the full files are available from \url{www.exomol.com}. The States file contains energy term values, state uncertainties, Land\'{e}-$g$ factors \citep{jt656}, lifetimes \citep{jt624} and quantum numbers. The Transition file contains Einstein A coefficients. The partition functions are also included as part the standard line list compilation.

\begin{table}
\centering
\caption{Line list statistics for each isotopologue of SiN.} 
\label{t:iso}
 \begin{tabular}{lrrr}
  \hline \hline
 Isotopologue & $g_{\rm ns}$ & $N_{\rm states}$ & $N_{\rm trans}$\\ 
 \hline
   \SiNm & 3 & 131935 & 43646806 \\
  $^{29}$Si$^{14}$N & 6 & 132335 & 43946969 \\
   $^{30}$Si$^{14}$N & 3 & 132706 & 44223730 \\
   $^{28}$Si$^{15}$N & 2 & 133460 & 44816182 \\
 \hline \hline
\end{tabular}

$g_{\rm ns}$:  Nuclear spin degeneracy; \\
$N_{\rm states}$: Number of states;\\
$N_{\rm trans}$:  Number of transitions. \\
\end{table}

The calculated energies were replaced with the MARVEL values (MARVELised), where available. We have used the labels 'Ca','EH'  and 'Ma' in the penultimate column of the States file to indicate if the energy value is calculated using \textsc{Duo}, derived using PGOPHER or MARVELised, respectively.  

The uncertainty values in the States file correspond to two cases: the MARVEL uncertainties are used for MARVELised energies, while for the calculated values the following approximate expression is used:
\begin{equation}
\label{e:unc}
{\rm unc} = a v + b J(J+1),
\end{equation}
where $a$ and $b$ are electronic state dependent constant, given in Table~\ref{t:unc:a:b}. For the \XS, \AS\ and \BS\ states uncertainties were estimated based on the progression of residuals  from Fig.~\ref{fig:residueAE} as average increases of obs.-calc. in $v$ and  $J$ for each state shown.

\begin{table*}
\centering
\caption{An extract from the states file of the \name\ line list for  \SiNm.}
\label{t:states}
{\tt  \begin{tabular}{rrrrrrrcclrrrrcr} \hline \hline
$i$ & Energy (\cm) & $g_i$ & $J$ & unc &  $\tau$ & $g$& \multicolumn{2}{c}{Parity} 	& State	& $v$	&${\Lambda}$ &	${\Sigma}$ & $\Omega$ & Label & Calc. \\ \hline
2  & 1138.471940 & 6 & 0.5 & 0.200000 & 0.5211 & 2.001919  & + & e & X2Sigma+ & 1 & 0 & 0.5  & 0.5 & Ma & 1138.405398 \\
3  & 2017.654928 & 6 & 0.5 & 0.010000 & 0.0015 & 0.000607  & + & e & A2Pi     & 0 & 1 & -0.5 & 0.5 & Ma & 2017.651598 \\
4  & 2263.632470 & 6 & 0.5 & 0.400000 & 0.1714 & 2.000967  & + & e & X2Sigma+ & 2 & 0 & 0.5  & 0.5 & Ma & 2263.834585 \\
5  & 3037.121068 & 6 & 0.5 & 0.006000 & 0.0006 & 0.000570  & + & e & A2Pi     & 1 & 1 & -0.5 & 0.5 & Ma & 3037.120364 \\
6  & 3375.983139 & 6 & 0.5 & 1.000000 & 0.0368 & 2.001071  & + & e & X2Sigma+ & 3 & 0 & 0.5  & 0.5 & Ma & 3376.117442 \\
7  & 4044.188981 & 6 & 0.5 & 0.200375 & 0.0004 & 0.000245  & + & e & A2Pi     & 2 & 1 & -0.5 & 0.5 & EH & 4044.208257 \\
8  & 4475.345990 & 6 & 0.5 & 1.600000 & 0.0124 & 2.001414  & + & e & X2Sigma+ & 4 & 0 & 0.5  & 0.5 & Ma & 4475.285785 \\
9  & 5038.850133 & 6 & 0.5 & 0.300375 & 0.0003 & -0.000102 & + & e & A2Pi     & 3 & 1 & -0.5 & 0.5 & EH & 5038.860507 \\
10 & 5561.363517 & 6 & 0.5 & 0.100750 & 0.0057 & 2.001737  & + & e & X2Sigma+ & 5 & 0 & 0.5  & 0.5 & Ca & 5561.363517 \\
11 & 6021.067289 & 6 & 0.5 & 0.400375 & 0.0002 & -0.000370 & + & e & A2Pi     & 4 & 1 & -0.5 & 0.5 & EH & 6021.017509 \\
12 & 6634.354317 & 6 & 0.5 & 0.120750 & 0.0032 & 2.001917  & + & e & X2Sigma+ & 6 & 0 & 0.5  & 0.5 & Ca & 6634.354317 \\
13 & 6990.593938 & 6 & 0.5 & 0.500375 & 0.0002 & -0.000488 & + & e & A2Pi     & 5 & 1 & -0.5 & 0.5 & Ca & 6990.593938 \\
14 & 7694.239922 & 6 & 0.5 & 0.140750 & 0.0020 & 2.001777  & + & e & X2Sigma+ & 7 & 0 & 0.5  & 0.5 & Ca & 7694.239922 \\
15 & 7947.534233 & 6 & 0.5 & 0.600375 & 0.0002 & -0.000299 & + & e & A2Pi     & 6 & 1 & -0.5 & 0.5 & Ca & 7947.534233 \\
16 & 8740.963173 & 6 & 0.5 & 0.160750 & 0.0014 & 2.000443  & + & e & X2Sigma+ & 8 & 0 & 0.5  & 0.5 & Ca & 8740.963173 \\
17 & 8891.787413 & 6 & 0.5 & 0.700375 & 0.0001 & 0.001068  & + & e & A2Pi     & 7 & 1 & -0.5 & 0.5 & Ca & 8891.787413 \\
\hline
\hline
\end{tabular}}
\mbox{}\\
{\flushleft
$i$:   State counting number.     \\
$\tilde{E}$: State energy term values in \cm. \\
$g_i$:  Total statistical weight, equal to ${g_{\rm ns}(2J + 1)}$.     \\
$J$: Total angular momentum.\\
unc: Uncertainty, \cm.\\

$\tau$: Lifetime (s$^{-1}$).\\
$g$: Land\'{e} $g$-factors. \\
$+/-$:   Total parity. \\
State: Electronic state.\\
$v$:   State vibrational quantum number. \\
$\Lambda$:  Projection of the electronic angular momentum. \\
$\Sigma$:   Projection of the electronic spin. \\
$\Omega$:   Projection of the total angular momentum, $\Omega=\Lambda+\Sigma$. \\
Label: 'Ma' is for MARVEL,'EH' is  PGOPHER generated and 'Ca' is for Calculated (using Duo). \\
}
\end{table*}

\begin{table}
\centering
\caption{An extract from the transitions file of the \name\ line list for  \SiNm.}
\label{t:trans}
\tt
\centering
\begin{tabular}{rrrr} \hline\hline
\multicolumn{1}{c}{$f$}	&	\multicolumn{1}{c}{$i$}	& \multicolumn{1}{c}{$A_{fi}$ (s$^{-1}$)}	&\multicolumn{1}{c}{$\tilde{\nu}_{fi}$} \\ \hline
34822  & 33867  & 6.3723E+01 & 4047.041414 \\
2808   & 2455   & 2.0314E-04 & 4047.041433 \\
63044  & 62682  & 1.4457E-06 & 4047.041573 \\
61850  & 60890  & 1.5342E-01 & 4047.042750 \\
65187  & 65424  & 1.1950E+02 & 4047.042900 \\
75707  & 75959  & 2.7837E+00 & 4047.043952 \\
37830  & 36872  & 9.1146E-02 & 4047.044029 \\
49826  & 49468  & 9.3588E-04 & 4047.044408 \\
18065  & 18306  & 6.1554E+01 & 4047.045440 \\
53117  & 53365  & 5.3547E-02 & 4047.046876 \\
70272  & 69910  & 5.1581E-03 & 4047.047010 \\
106232 & 106467 & 5.6296E-01 & 4047.047815 \\
 \hline\hline
\end{tabular} \\ \vspace{2mm}
\rm
\noindent
$f$: Upper  state counting number;\\
$i$:  Lower  state counting number; \\
$A_{fi}$:  Einstein-$A$ coefficient in s$^{-1}$; \\
$\tilde{\nu}_{fi}$: transition wavenumber in \cm.\\
\end{table}

\begin{table}
\centering
\caption{$a$ and $b$ constants (\cm) defining state dependent uncertainties via Eq. \eqref{e:unc}.} 
\label{t:unc:a:b}
 \begin{tabular}{llc}
  \hline \hline
 State & $a$ & $b$ \\ 
 \hline
 \XS   &  0.02  & 0.001  \\
 \AS   &  0.1  & 0.001  \\
 \BS   &  0.5  & 0.001  \\
 All other   &  0.8  & 0.001  \\
 \hline \hline
\end{tabular}
\end{table}

\subsection{Overall Spectra}

\begin{figure}
    \centering
    \includegraphics[width=0.80\textwidth]{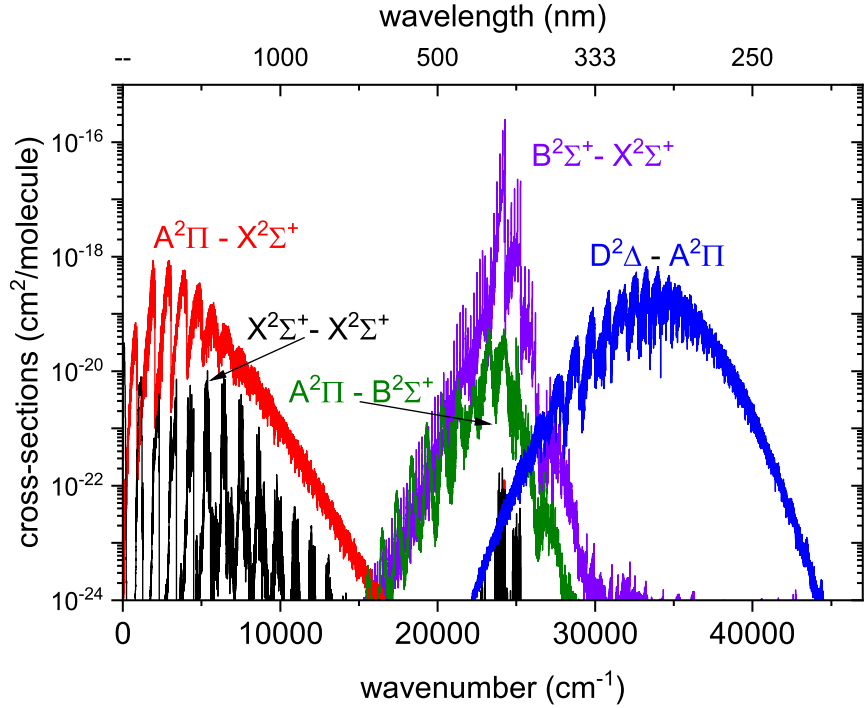}
    \caption{Simulated absorption spectrum of SiN at 2000 K showing the main bands of the system. A Gaussian profile of the half width of half maximum (HWHM) of 1~\cm\ was used. }
    \label{fig:Bands}
\end{figure}

\begin{figure}
    \centering
    \includegraphics[width=0.80\textwidth]{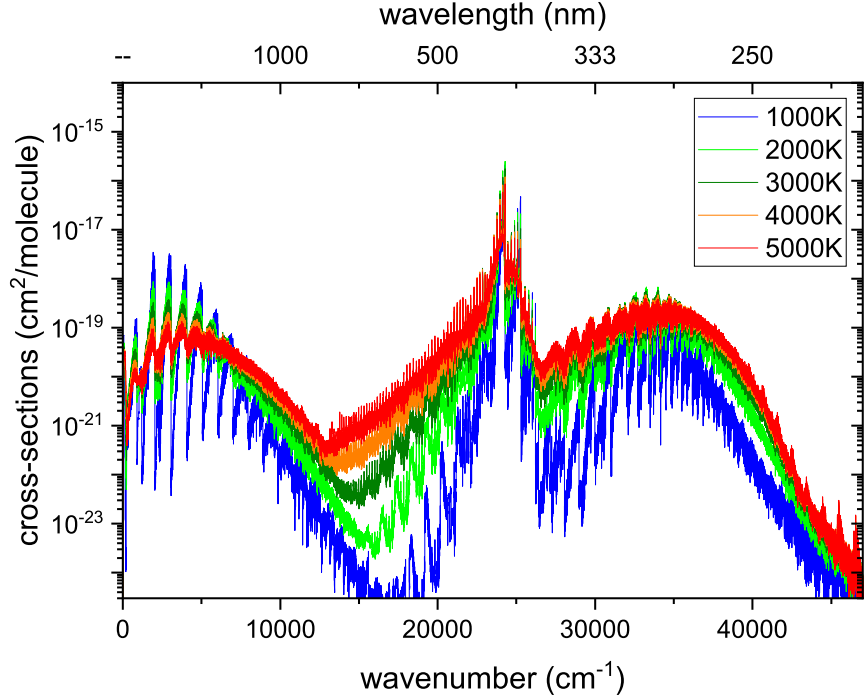}
    \caption{Temperature dependence of the SiN spectra using the \name\ line list. A Gaussian profile with HWHM of 1~\cm\ was used. }
    \label{fig:temperatureSpectra}
\end{figure}

To demonstrate the accuracy of the \name\ line list, several spectra were calculated, analysed and compared to available laboratory measurements. Figure~\ref{fig:Bands} illustrates the main bands of SiN at 2000 K. The dominance of the \XS--\AS\ in the 0--5000 \cm\ region confirms  why it is  hard to detect the comparatively weak \XS--\XS\ transitions. Additionally, while the \XS--\BS\ band system appears to be overall stronger than \AS-- \BS, the latter should still be detectable in the 18000 -- 24000 \cm\ region according to our model, in line with observations of these vibronic bands by \citet{06OjGoxx.SiN} and \cite{85FoLuAm.SiN}. Additionally, in the region above 28000 \cm\ the \AS\ -\DS\ band becomes dominant, which agrees with previous vibronic observations in this region  by \citet{76BrDuHo.SiN}.
Figure \ref{fig:temperatureSpectra} shows how the spectrum of \SiNm\ changes with increasing temperature.

\subsection{\BS--\XS\ band}

\begin{figure}
    \centering
    \includegraphics[width=0.44\textwidth]{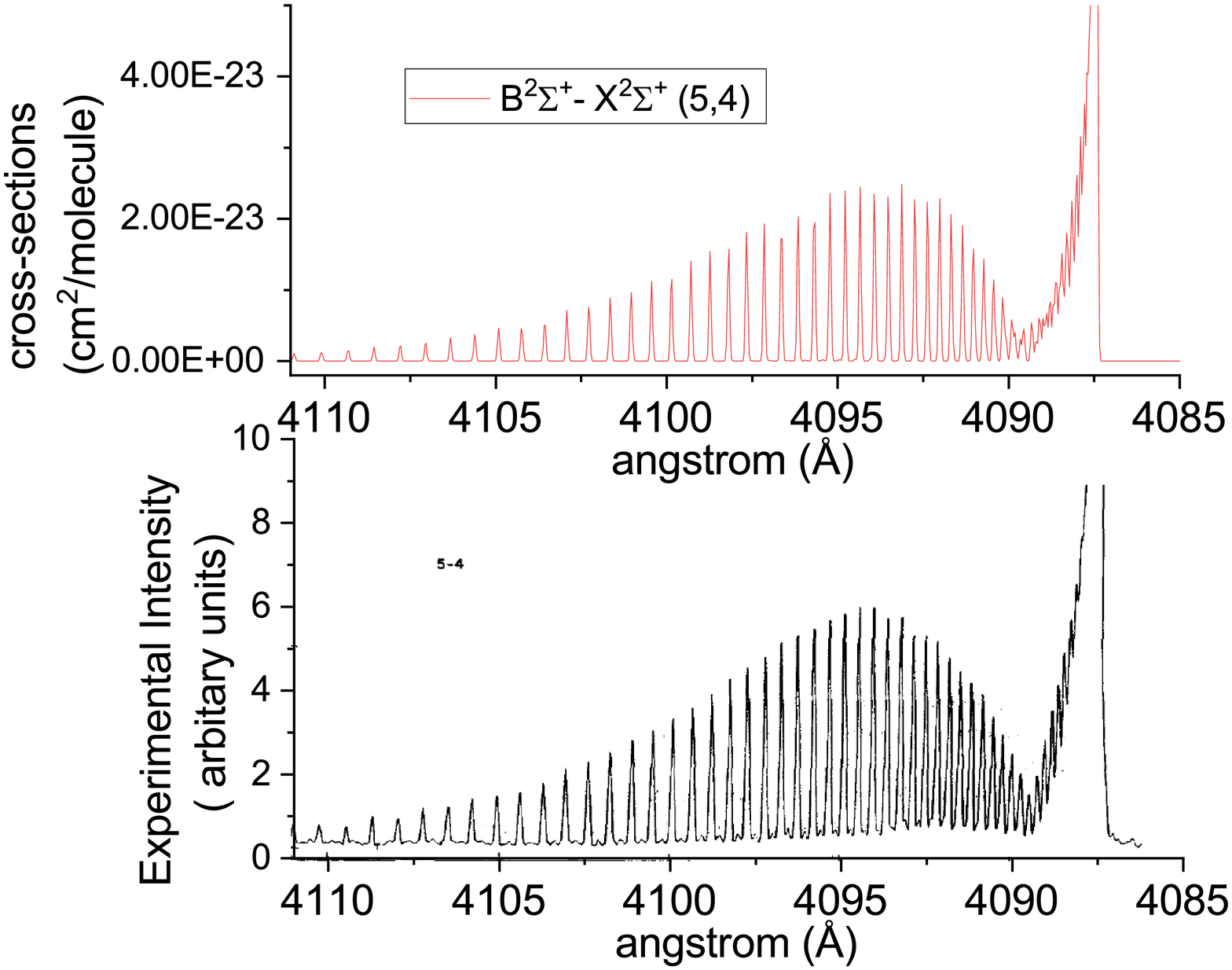}
    \includegraphics[width=0.44\textwidth]{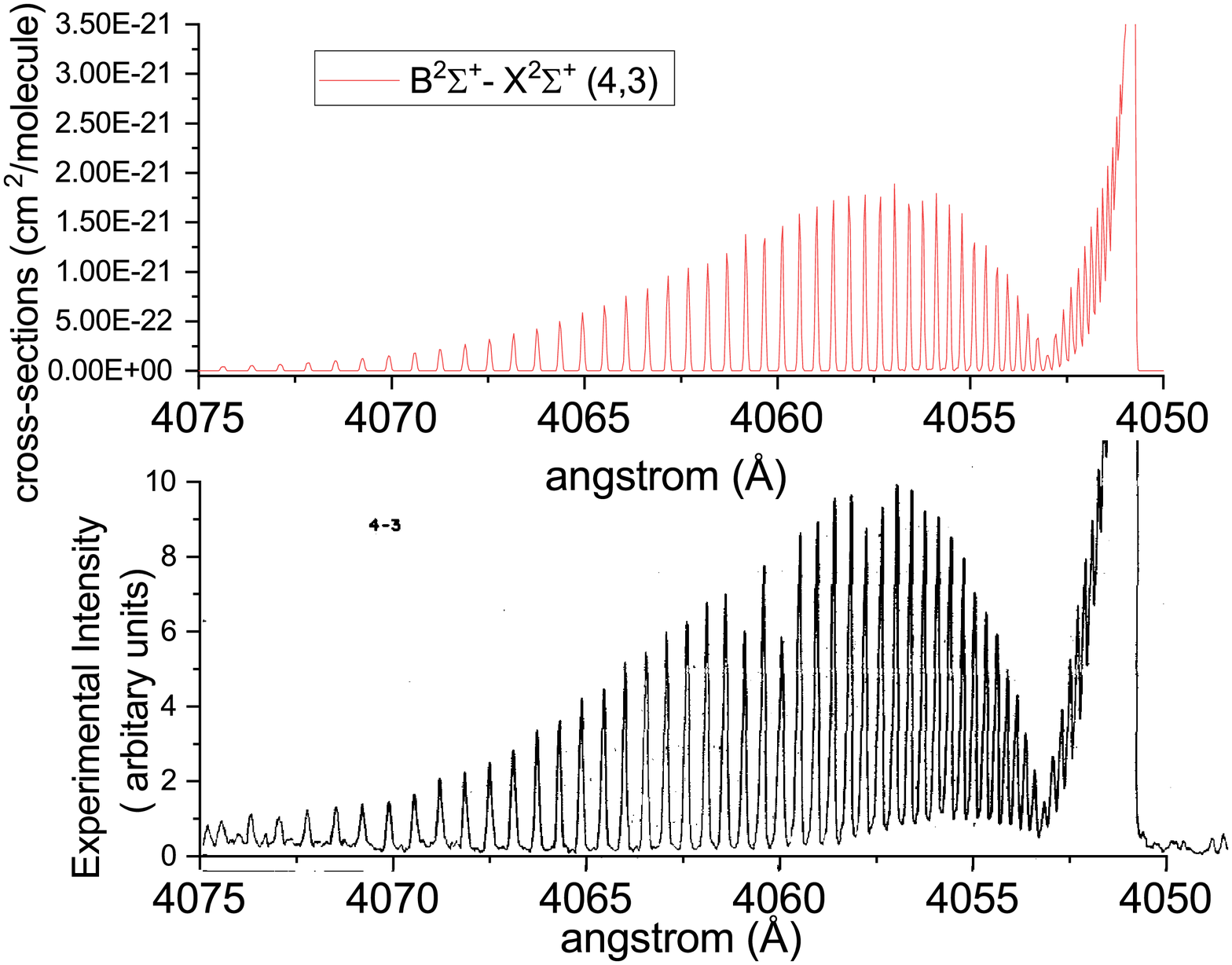}
    \caption{Simulated absorption spectra of SiN showing the \BS--\XS\ (5,4) and (4,3) bands at temperature of 392~K and 412~K respectively. A comparison to the experimental spectra from \citet{65ScBrxx.SiN} is provided. A Gaussian profile HWHM of 0.2~\cm\ was used.}
    \label{fig:65ScBr}
\end{figure}

\BS--\XS\ is the strongest electronic band in the system, with the largest number of experimental observations due to it being the easiest to detect. In Figure \ref{fig:65ScBr} we simulate the \BS--\XS\ (5,4) and (4,3) band at rotational temperature of 392~K and 412~K respectively to provide direct comparison with the experiment. The simulated spectra is also adjusted to be in air rather than in vacuum to align with the experimental spectra. This is achieved by using the IAU standard of conversion adopted by \citet{91Morton}:
\begin{equation}\label{e:conversion}
\lambda_{\rm air}= \frac{\lambda_{\rm vacuum}} {(1.0 + 2.735182\times10^{-4} + \frac{131.4182}{\lambda_{\rm vac}^{2}} + \frac{2.76249\times^{8}}{\lambda_{\rm vac}^{4}})}
\end{equation}
where $\lambda_{\rm air}$ and $\lambda_{\rm vac}$ are wavelengths in air and vacuum respectively. The overall agreement on the rotational structure is within the uncertainty provided.

\subsection{\AS--\XS\ band}

\begin{figure}
    \centering
    \includegraphics[width=0.8\textwidth]{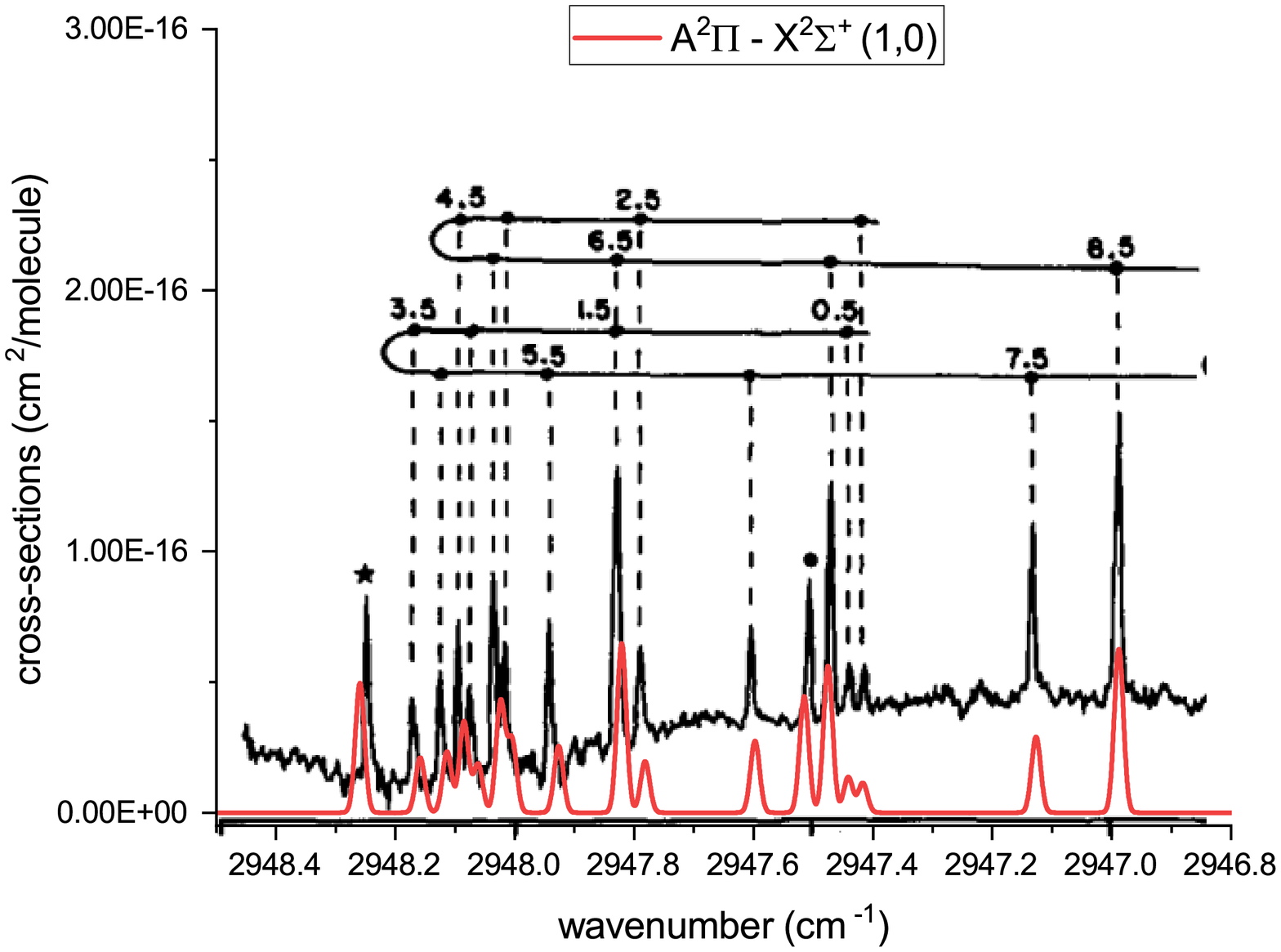}
    \caption{Simulated absorption spectrum of SiN at effective temperature of 740~K showing the \AS--\XS\ (1,0) band. A Gaussian profile of HWHM of 0.009 ~\cm\ was used. A comparison with the experimental spectra from \citet{85FoLuAm.SiN} is provided, with calculated spectra in red and experimental in black.}
    \label{fig:85FoLuAm}
\end{figure}

The \AS--\XS\ band system was first detected by \citet{13Jexxxx.SiN}, however because of the vicinity of  the \AS\ and \XS\ states it has proved difficult to study experimentally.
Figure \ref{fig:85FoLuAm} shows the comparison of our simulated spectra of the \AS--\XS\ (1,0) band with the experiment of \citet{85FoLuAm.SiN}. The spectrum is simulated at the effective temperature of 740~K. The position of the bands agrees to 0.05~\cm\ which is within our calculated uncertainties. 


\begin{figure}
    \centering
    \includegraphics[width=0.8\textwidth]{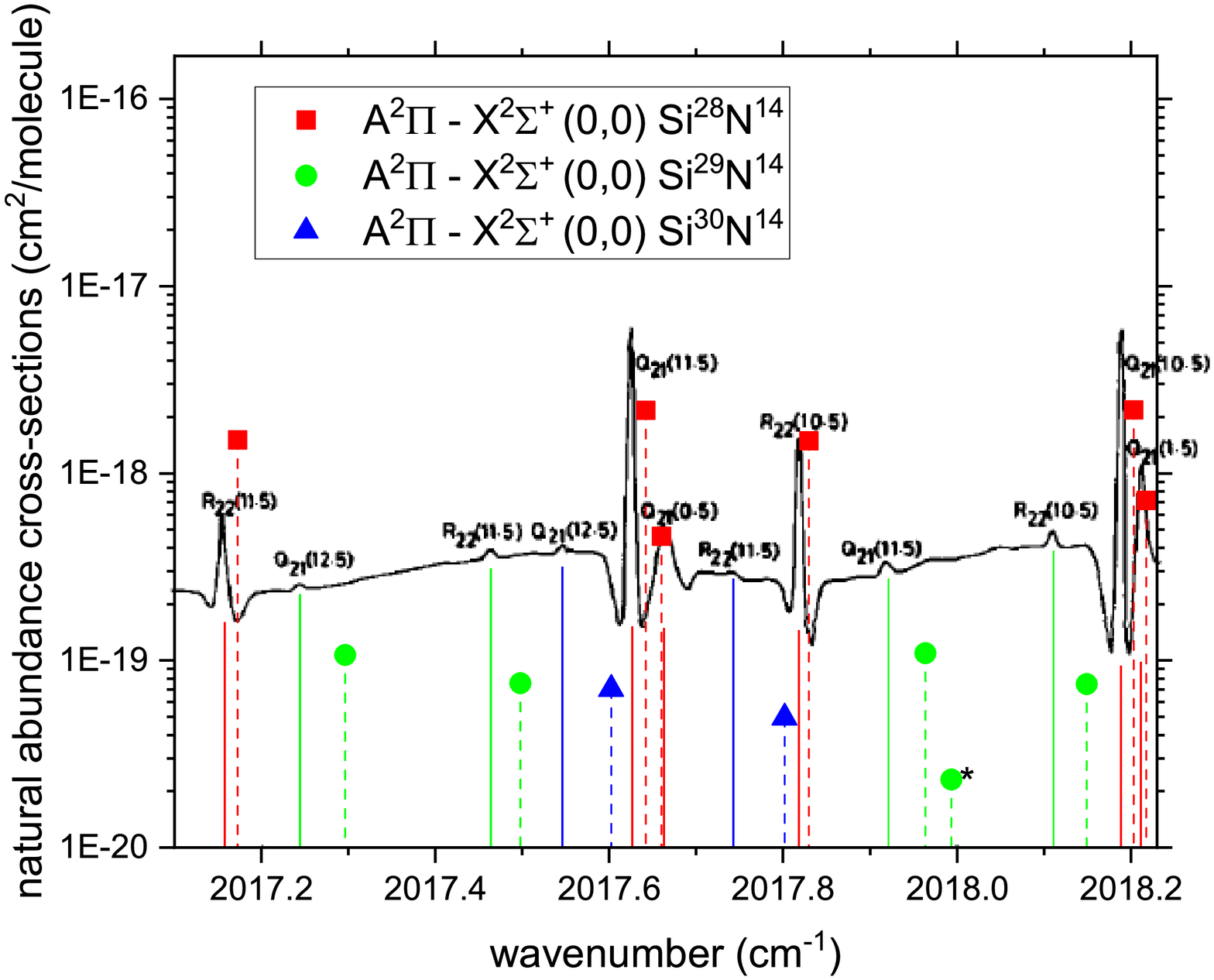}
    \caption{Comparison of the simulated stick spectrum and experimental emission spectrum of the SiN isotpologues (\SiNm, $^{29}$Si$^{14}$N, $^{30}$Si$^{14}$N) at 300~K. Straight lines are experimentally observed line positions and dashed lines with points represent calculated. The asterisk indicates the Q$_{22}$(0.5) line for $^{29}$Si$^{14}$N not observed experimentally by \citep{88YaHiYa.SiN}. }
    \label{fig:88YaHiYaIso}
\end{figure}

Figure \ref{fig:88YaHiYaIso} shows a comparison of the simulated spectra (shown as sticks) for different silicon isotopes of the SiN molecule with the experimental spectra of \citet{88YaHiYa.SiN}. The overall agreement for  \SiNm is $\sim${}0.016~\cm{}, for $^{29}$Si$^{14}$N is $\sim${}0.035~\cm{} and for $^{30}$Si$^{14}$N is $\sim${}0.055~\cm{}. The intensities were adjusted based on the natural abundance for each silicon isotope. These are defined as 92.2\%, 4.68\% and 3.09\%  for \SiNm\ ,$^{29}$Si$^{14}$N, $^{30}$Si$^{14}$N respectively. Additionally, in the same range a Q$_{22}$(0.5) line of   \AS--\XS\ (0,0) for $^{29}$Si$^{14}$N should be present, but it was probably too weak to be observed during the experiment. This line is indicated with the an asterisk.

\subsection{Lifetimes}

There are few precise experimental measurements of the lifetimes of different electronic states of SiN in the literature. \citet{84WaAvDr.SiN} report the \BS\ $v=1$ vibrational state to have a lifetime of  200~ns $\pm 10$~ns ($T_{\rm rot }$ = 500~K). Our calculated value for this vibronic state is lower: 130~ns ($J=0.5$), slowly increasing to $ 150$~ns  for $J=100$~ns and even 200~ns for $J=160.5$ ($v=1$, \BS). Additionally there have been several theoretical works \citep{18XiShSu.SiN, 15KaBaxx.SiN} which showed that for the \AS\ state the lifetimes decrease with the vibrational levels as lower levels are much less energetically accessible due to the proximity of the \XS\ state. This can be seen in more detail in Table \ref{tab:lifetimes}, where we compare our calculated lifetimes with those of \citet{18XiShSu.SiN}; the agreement is good.

\begin{table*}
\caption{Comparison of lifetimes from our current work (A) and \citet{18XiShSu.SiN} (B). An experimental lifetime 200 $\pm$ 10~ns  was reported by  \citet{84WaAvDr.SiN} for $v'=1$ of \BS.}
\begin{tabular}{rrrrr}
    \hline
    \hline
   & \multicolumn{2}{c}{\AS / ~$\mu$s}   & \multicolumn{2}{c}{\BS / ~ns }  \\
      \hline
$v'$ & A & B & A & B  \\
    \hline
       0  &         1460  &         1660  &            124   &             129  \\
       1  &          610  &          679  &            130   &             136  \\
       2  &          385  &          429  &            137   &             143  \\
       3  &          281  &          312  &            143   &             150  \\
       4  &          221  &          245  &            150   &             157  \\
       5  &          182  &          202  &            156   &             164  \\
       6  &          155  &          172  &            163   &             170  \\
       7  &          135  &          149  &            169   &             176  \\
       8  &          121  &          132  &            175   &             181  \\
       9  &          109  &          119  &            181   &             187  \\
      10  &           98.3  &          109  &            186   &             192  \\
      11  &           90.5  &           99.8  &            192   &             196  \\
      12  &           84.0  &           92.6  &            197   &             201  \\
      13  &           78.5  &           86.6  &            202   &             205  \\
      14  &           73.9  &           81.5  &            207   &             210  \\
    \hline
    \hline
\end{tabular}

\label{tab:lifetimes}
\end{table*}

\subsection{Partition function}

\begin{figure}
    \centering
    \includegraphics[width=0.7\columnwidth]{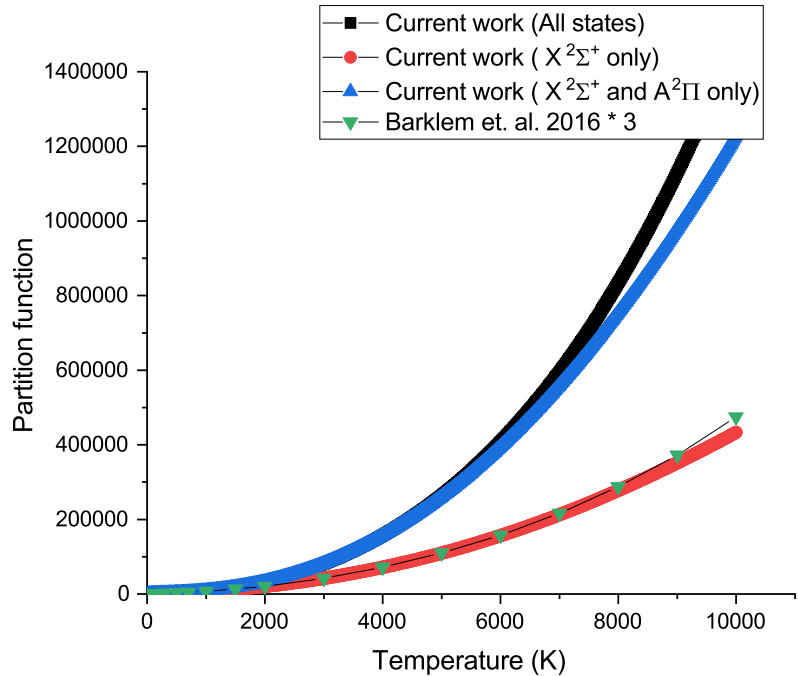}
    \caption{Comparison of the partition functions for $^{28}$Si$^{14}$N: values from this work and the work of \citet{16BaCoxx.partfunc}.}
    \label{f:partition_fucntion}
\end{figure}

For the partition function of \SiNm, we  follow the ExoMol and HITRAN \citep{jt692} convention and include the full nuclear spin. This means that our convention is different to that of recently reported by \citet{16BaCoxx.partfunc} with whom we compare our results in Fig.~\ref{f:partition_fucntion}. In order for comparison to be in equivalent convention, their partition function was multiplied by the factor of three as $^{14}$N has a nuclear spin degeneracy of 3 and $^{28}$S has nuclear spin degeneracy of 0. The differences at higher temperatures can be attributed to incompleteness of the model used by \citet{16BaCoxx.partfunc}, which is demonstrated by comparing partition functions by using only \XS\ state, \XS\ and \AS\ states and all states: the partition function of  \citet{16BaCoxx.partfunc} appears to be based on the  \XS\ state only.  As evident from Fig.~\ref{f:partition_fucntion}, the contribution from the \AS\ state cannot be neglected  due to its low excitation energy.

Partition functions in for $T= 1 - 3000$ K
in steps of 1~K are available the 4
isotopologue, $^{28}$Si$^{14}$N, $^{29}$Si$^{14}$N, $^{30}$Si$^{14}$N and $^{28}$Si$^{15}$N,
are available via the ExoMol website.

\section{Conclusion}
New IR and UV line lists called \name\ for isotopologues of SiN are presented. \name\ is available from \url{www.exomol.com} \citep{jt810} and from \url{www.zenodo.org} \citep{Zenodo}.  As part of the line list construction, a MARVEL analysis for \SiNm\ was performed. All experimental line positions from the literature (to the best of our knowledge) currently available for the $X$--$X$, $A$--$X$ and $B$--$X$ systems were processed to generate a MARVEL  set of empirical energies of SiN.  An accurate spectroscopic model for SiN was built: \ai\ (T)DMC and empirically refined PECs, SOCs, EAMCs, using both previously derived MARVEL energies as well as PGOPHER generated energies.

The line list was MARVELised, where the theoretical energies are replaced  with the MARVEL  values (where available). The  line list provides uncertainties of the rovibronic states in order to help in high-resolution applications.  Comparisons of simulated spectra with experiment for both $A$--$X$ and $B$--$X$ system show close agreement. Additionally lifetimes, Land\'{e}-g factors and partition functions are provided. The importance of inclusion of the energies of the excited state \AS\ when computing the partition function of SiN is demonstrated.

\section*{Acknowledgements}

  This work was supported by UK STFC under grant ST/R000476/1 and the European Research Council (ERC) under the European Union’s Horizon 2020  research and innovation programme through Advance Grant number  883830.

\section{Data Availability} 

The data underlying this article are available in the article and in its online supplementary material, including (1)  the \textsc{Duo} input files for each isotopologue containing all the potential energy, (transition) dipole moment and coupling curves of SiN used in this work, (2) the MARVEL data set, (3) digitalised line data from \citet{75Lixxxx.SiN}, (4) extracted ab initio curves as part of the excel, (5) PGOPHER file used to generate energy levels for two low lying states, (6) the temperature-dependent partition function of \SiNm\ up to 10000~K. The  SiNful line lists for  $^{28}$Si$^{14}$N, $^{29}$Si$^{14}$N, $^{30}$Si$^{14}$N and $^{28}$Si$^{15}$N  are  available from \href{www.exomol.com}{www.exomol.com}. 


\bibliographystyle{mnras}

\label{lastpage}

\end{document}